%% file: latex/acl_latex.tex
\pdfoutput=1
\documentclass[11pt]{article}

\usepackage[preprint]{acl}

\usepackage{times}
\usepackage{latexsym}

\usepackage[T1]{fontenc}
\usepackage[utf8]{inputenc}

\usepackage{microtype}

\usepackage{inconsolata}
\usepackage{todonotes}
\usepackage{graphicx}
\usepackage{graphicx}
\usepackage{amsmath}
\usepackage{multirow}
\usepackage{hyperref}
\usepackage{xurl}
\usepackage{url}

\definecolor{red1}{rgb}{0.9,0.1,0.1}
\definecolor{red2}{rgb}{0.6,0.4,0.4}
\definecolor{grn1}{rgb}{0.1,0.9,0.1}
\definecolor{grn2}{rgb}{0.4,0.6,0.1}
\definecolor{grey}{rgb}{0.5,0.5,0.5}

\usepackage{subcaption}
\title{Educators' Perceptions of Large Language Models as Tutors: \\Comparing Human and AI Tutors in a Blind Text-only Setting}

\author{Sankalan Pal Chowdhury\thanks{For queries contact \texttt{spalchowd@ethz.ch}} \\
  ETH Zurich
  \\\And
  Terry Jingchen Zhang\\
  ETH Zurich
  \\\And
  Donya Rooein\\
Bocconi University
  \\\AND
  Dirk Hovy\\
Bocconi University
  \\\And
  Tanja K\"aser\\
  EPFL
  \\\And
  Mrinmaya Sachan\\
  ETH Zurich
  }

\begin{document}
\maketitle
\begin{abstract}
% The rapid development of Large Language Models  (LLMs) opens up the possibility of using them %in new domains, such as using them 
% as personal tutors. This possibility has led to the development of several text-based intelligent tutoring systems and learning assistants using LLMs as back-ends with various degrees of engineering. %Today, most of these tools communicate with learners primarily in a text-only setting.
% % problem statement. Something like: 
% However, successful tutoring involves (social) skills beyond knowledge transfer that LLMs might lack. 
% In this study, we compare human tutors with LLM tutors in a text-only setting, in terms of various desirable qualities of a tutor. We ask other human tutors to compare the performance of an AI tutor with that of a human tutor in teaching grade-school math word problems on various such qualities. We find that our subjects perceive LLMs to be outperforming human tutors in terms of being concise and scaffolding quality of their responses, while maintaining engagement and empathy. Our study paints a positive picture of LLMs as tutors and indicates the possibility that these models can be used to reduce load on human teachers in the future as their multimodal capabilities in tutoring improve over time.

% Donya abstract
The rapid development of Large Language Models (LLMs) opens up the possibility of using them as personal tutors. This has led to the development of several %text-based 
intelligent tutoring systems and learning assistants that use LLMs as back-ends with various degrees of engineering. In this study, we seek to compare human tutors with LLM tutors %in a text setting 
in terms of engagement, empathy, scaffolding, and conciseness. We ask human tutors to annotate and compare the performance of an LLM tutor with that of a human tutor in teaching grade-school math word problems on these qualities. We find that annotators with teaching experience perceive LLMs as showing higher performance than human tutors in all 4 metrics. The biggest advantage is in empathy, where 80\% of our annotators prefer the LLM tutor more often than the human tutors. Our study paints a positive picture of LLMs as tutors and indicates that these models can be used to reduce the load on human teachers in the future.

\end{abstract}

\section{Introduction}
%\mrinmaya{I guess we will rephrase this}
Recent improvements in Large Language Models (LLMs) have opened up the possibility of using them in multiple new domains, including as personal tutors. This possibility has led to the development of several Intelligent Tutoring Systems (ITSs) and learning assistants  \cite{schmucker2023ruffle,liffiton2023codehelp,lieb2024student,selfcite1} that use LLMs as backends with various degrees of engineering. %At the same time, s
Surveys by \texttt{Intelligent.com}  \cite{Intelligent} and \texttt{DEC Singapore} \cite{DECReport} indicate that a large number of students are already using LLMs like ChatGPT in educational roles such as tutoring.
%\mrinmaya{Can we macro things like ChatGPT, MWPTutor, MathDial...? }%Maybe intelligent.com and DEC Singapore can be texttt}

Despite their popularity, a clear understanding of the pedagogical effectiveness of educational chatbots, especially compared to humans, is lacking. The common way of using LLMs as tutor is to interact with them via a chat interface, where the LLM roleplays a tutor. It is known that the full benefit of a human tutor goes well beyond verbal or textual communication \cite{bambaeeroo2017impact}, giving human tutors an advantage over LLM-based tutors. However, it remains unclear how LLM-based tutors compare with their human counterparts, in this chat setting.  %even within the limited setting of text-only tutoring, where their main strength lies. 
A good tutor keeps students \textbf{engaged}, \textbf{empathises} with their struggles, \textbf{scaffolds} them to correct answers, all while keeping the conversation to the point and \textbf{concise}. Is an LLM-based tutor capable of doing the same?
% \mrinmaya{It is also debatable if we are looking at the entirety of text-only tutoring. Maybe we can further refine it to say chat-based text-only tutoring. Also state that we do this in the context of grade-school math word problems somewhere.}

In this study, we compare human tutors with LLM-based tutors, through the dialogs generated via chat interfaces. Our main research question is:
\begin{quote}
    \textit{How do LLM-based tutors compare to %average?
    human tutors in terms of 
    engagement, empathy, scaffolding, and conciseness?}% for Math Word Problems (MWPs)\footnote{The domain is mostly decided due to dataset availability}}
\end{quote}
Although there have been some recent attempts to compare learning gains from LLM-based tutors and human tutors (see Sec \ref{sec:Related}), these studies focus on the observable outcomes of learning gains. Our study seeks to complement these studies by instead focusing only on the latent factors (we will provide a more detailed definition and justification in Sec. \ref{sec:metrics}), and run comparisons on these directly. We believe that knowing how LLMs stand on these would allow researchers to better focus on what to improve in these models.

Our contributions are:
\begin{enumerate}
    \item We create a setup to ask human annotators to compare tutoring dialog snippets in a blind pairwise preference selection setting.
    \item We use this setup to have teachers compare a human tutor with an LLM tutor on a dataset of MWPs to identify how they compare the $4$ latent factors involved in student learning.
    \item We publicly release the \href{https://polybox.ethz.ch/index.php/s/nSF92x5wEjydaDa}{annotation data} consisting of 210 annotated dialog pairs to help future research better align LLM outputs to human judgments.  
\end{enumerate}
% \donya{Here we need to mention the RQs and contribution in a more clear way.}\\
%Our study compares two tutoring datasets working on the same set of Math Word Problems (MWPs). Solving MWPs required a combination of linguistic and mathematical reasoning, and thus have been one of the gold standards for evaluating LLMs. 

Our experiments find that annotators with teaching experience perceive the LLM tutor to be more engaging and empathetic while also being concise and better at scaffolding the student. This also aligns with LLMs self-judgments, though fine-grained tendencies are quite different.
\section{Related Work}
\label{sec:Related}
%\vspace{-3mm}
% \subsection{Evaluating Human Tutors}
% \donya{AI-based or LLM-based tutor}
\subsection{Designing and Evaluating LLM Based Tutors}
% copied from Arxiv version of the Italy Paper
With the recent progress in LLMs, there have been several efforts to develop and evaluate LLM-based tutors. A large number of these have focused on computer science and programming education \cite{cref,qi2024knowledgecomponentbasedmethodologyevaluatingai,liffiton2023codehelp,Kazemitabaar_2024,Jacobs_2024,10.1145/3626252.3630938,Lyu_2024,li2024tutorlyturningprogrammingvideos,choudhuri2023farwetriumphstrials,Pankiewicz_2024}, but there have also been developments in domains like mathematics  \cite{Chowdhury2024AutoTutorML,10297824,10.1371/journal.pone.0304013}, language learning  \cite{doi:10.1080/2331186X.2024.2355385,Park_2024,selfcite1}, health sciences  \cite{info:doi/10.2196/51344,chheang2024anatomyeducationgenerativeaibased,wang2024patientpsiusinglargelanguage} and other domains  \cite{thway2024battlingbotpoopusinggenai,chen2024effectiveness}. However, most of these works focus on the engineering behind developing the tutor, and if any evaluation is done, it is either in terms of learning gains, or in the terms of student self-reports of efficacy and motivation. Moreover, the comparison in these studies is always between having an LLM-based tutor and not having anything, and not with human tutors. Finally, we also lack an understanding of the factors contributing to a good quality tutor. %\mrinmaya{Can work on refining this. It suffices to say many papers do not contain learning experiments, and those that do, the evauations only focus on learning gains. And we lack an understanding of the factors contributing to a good quality tutor.}
%\vspace{-4mm}
\subsection{Comparing AI Tutors with Human Tutors}
Tutoring was established as one of the best ways to improve learning outcomes by Bloom in 1984 \cite{bloom2sigma}, and matching the learning gains of a human tutor has been one of the main targets of computer-based tutors ever since \cite{sleeman1982intelligent}. Several studies have compared the learning gains from different types of computer-based tutors with humans \cite{KULIK199175,WOS:A1995TQ89500002,VanLEHN01102011} and with the development of LLM-based tutors, the same has also been extended to LLM based tutors \cite{schmucker2023ruffle,zhang2024investigationeffectivenessapplyingchatgpt}. However, these works focus only on the final learning gain, not on the latent qualities that could cause it.

Since several computer-based tutors communicate in natural language, another line of work follows from Alan Turing's Imitation Game \cite{imitation}, which was later adapted into the `Bystander Turing Test' \cite{Bystander}. While our work is similar to this in terms of the text-only setup and blind selection, we differ in that instead of asking the annotator to determine which party is human, we ask them to determine which one is better on a set of metrics.
%\vspace{-3mm}
\section{Method}
%\vspace{-2mm}
\subsection{Datasets}
To compare human and LLM tutors, we need 
%datasets that contain one-on-one textual conversation between a (human) tutor and a student. We would then need to obtain a parallel corpus with everything else held the same, only the teacher replaced by a LLM tutor to compare.
parallel data sets of student-tutor text interactions, both for human and LLM tutors.
Among the limited one-on-one tutoring datasets available, we cannot use data sets such as TSCC \cite{caines-etal-2020-teacher} or CIMA \cite{stasaski-etal-2020-cima} because they use human students, and a fair comparison would require us to repeat the LLM side of the experiment with the same humans. To avoid doing this, we draw our conversations from MathDial \cite{macina-etal-2023-mathdial} for the human side because the students in this dataset are simulated by AI.

MathDial consists of about $3000$ tutor-student conversations fixing student errors on MWPs. The MWPs were sampled from the GSM8K dataset \cite{cobbe2021trainingverifierssolvemath}, while the misconceptions were generated using InstructGPT. The authors hired annotators with teaching experience on Prolific to converse with an InstructGPT instance pretending to be a student having the particular misconception, a setup we can easily replicate at little cost. The annotators were prescribed some pedagogical suggestions urging them to avoid giving out answers directly but were otherwise encouraged to behave as they would when tutoring a real student.

Moving on to the LLM side, we could simply use a modern LLM like GPT to repeat the conversations from MathDial with identical settings. However, this only works if we can ensure that the GPT model would never give incorrect feedback, for example stating that a student's answer is right when it is not. If a tutor has a chance of giving out wrong information, comparing its softer qualities is moot. Unfortunately, previous work has found that GPT4-turbo does make such mistakes \cite{Chowdhury2024AutoTutorML} and we found in our explorations that this is still the case for GPT4o, with $6$ out of the $30$ problems investigated having some issue ($1$ case where the teacher gave a wrong answer, $5$ cases of teacher telling the right answer but not verifying if the student agreed, including $2$ where the final teacher utterance included nonsensical phrases). Therefore, we instead use conversations from MWPTutor \cite{Chowdhury2024AutoTutorML}, a tutor based on LLMs which ensures correctness by imposing guardrails on top of GPT. 

MWPTutor uses a finite state transducer to prompt an LLM to generate the best teacher utterances and uses the same InstructGPT student model as MathDial. The paper proposes multiple versions of their system, but in this work, we make use of $\text{MWPTutor}^\text{live}_\text{GPT4}$  as it is the best according to their own metrics. Although MathDial consists of about $3000$ conversations, many of them repeat the same MWPs and incorrect solutions. Since MWPTutor only makes use of these two components, we restrict our study to one conversation per MWP. As such, we choose $210$ MWPs including all $45$ coming from the GSM8K test set. For MathDial, we pick the first conversation when sorted by timestamp. For MWPTutor, we use the conversations published by the authors for the test set MWPs, while for the remaining, we generate conversations using their publicly available code.

Note that despite the accuracy issue, we did perform a smaller study with GPT4o instead of MWPTutor, and found that the trends were not much different  (See Appendix \ref{sec:mdvgpt} for details).

% \donya{We need to clarify why we chose these two datasets and describe the characteristics we should consider if we decide to add more datasets.}
% We draw our conversations from MathDial \cite{macina2023mathdial} and MWPTutor \cite{10.1145/3657604.3662041} for Human and AI Tutors respectively. 

% MWPTutor is a later work that draws inspiration from Intelligent Tutoring Systems literature to replace the human in the scenario used by MathDial with an AI. It uses a Finite State Transducer to prompt an LLM to generate the best teacher utterances. The paper proposes multiple versions of their system, but in this work we make use of the $\text{MWPTutor}^\text{live}_\text{GPT4}$  as it is the best by their own metrics. 

% While MathDial consists of about $3000$ conversations, many of them repeat the same MWPs. Since MWPTutor only makes use of these two components, we restrict ourselves to one conversation per MWP. As such, we choose $210$ MWPs including all $45$ coming from the GSM8K test set. For MathDial, we pick the first conversation when sorted by timestamp. For MWPTutor, we use the conversations published by the authors for the test set MWPs, while for the remaining we generate conversations using their publicly available code.
%\vspace{-3mm}
\subsection{Metrics}
\label{sec:metrics}
Tutoring is a rather complex task, and thus it is hard to list desirable tutoring qualities that can be considered universally applicable.  The primary desiderata for our study are that we need a small set of metrics (so as to be able to evaluate them in a reasonable budget), which can be judged from text and are subjective enough to facilitate the comparison of two conversations. To obtain such a set of metrics, we drew inspiration from 3 main works. 

Ross, in his book \cite{macdonald2000master} identifies 6 goals for tutors, although this includes more administrative duties such as ``provide student perspective on school success''.  Walker \cite{walker2008twelve} surveyed several teachers in training, and identified $12$ desirable characteristics of teachers. Although too numerous and often requiring actions beyond a text-only setting, they serve as a good starting point for us. Maurya et al \cite{maurya2024unifying} unified several recent works to identify 8 metrics relevant to AI tutors. However, these metrics are often too precise, making it difficult to rank two conversations based on them. 

Inspired by these and other works mentioned in the definitions, we decided on four metrics to evaluate, which we discuss below. An important thing to point out here is that though these metrics have scientific grounding, they are all quite subjective, which means in certain cases choosing the better of a pair of conversations might become a matter of personal preference. Although we did not evaluate the original metrics in the above work with humans, we did run them through GPT, and the results are provided in the Appendix (see Section \ref{sec:other-models}). We also provide a full mapping between the metrics in the three aforementioned papers and our metrics in Section \ref{sec:correspond}

\noindent\textbf{Engagement:} Student engagement can be defined as `how involved or interested students appear to be in their learning'  \cite{EngagementDefinition}. All of Walker, Ross and Maurya  (see table \ref{tab:Corr}) use metrics that map to engagement. High student engagement is positively correlated with student learning outcomes  \cite{EngagementClassic}, and this effect has also been observed in recent studies on LLM tutors \cite{altememy2023ai,selfcite1}. 
\\
% Our definition of Engagement covers the desiderata of \textit{prepared} and  \textit{creative} from  \cite{walker2008twelve}, \textit{coherence} from  \cite{maurya2024unifying} and \textit{promote}
%It is therefore important for a tutor, AI or Human, to keep the student engaged in the conversation. We note at this point that in the limited scope of short text-only conversation, engagement is very similar to conversational uptake \cite{demszky2021measuring}.
\noindent\textbf{Empathy:}
Empathy is the ability of a tutor to understand the hardships a student is facing and to react in a way that keeps up their motivation. Empathy is seen as important in a teacher by most educators \cite{Empathy1,Empathy2}, and practical studies show that teachers' empathy is correlated with positive learning outcomes for at least some groups of students  \cite{bostic2014teacher,d2013autotutor}. Walker identifies multiple dimensions of empathy as essential, while Maurya and Ross also consider it important (see table \ref{tab:Corr}). One important thing to note here is that empathy in general is a rather broad term, and is often split into subcategories of emotional and cognitive empathy \cite{smith2006cognitive}. In this work, `Empathy' primarily refers to Emotional Empathy, whereas Cognitive Empathy is somewhat subsumed by Engagement. 

\noindent\textbf{Scaffolding:}
Scaffolding is the idea that a tutor should help a student succeed in a problem, not by directly revealing the answer, but by controlling elements of the problem solving process to enable the student to achieve the solution by themselves  \cite{wood1976role}. Doing so helps students to not just understand the solution of the problem at hand but also learn the concepts behind the solution, enabling them to solve similar problems thereafter. The first five metrics from Maurya all reflect forms of scaffolding, while Ross covers it with `promote independent learning' and `facilitate tutee insights into learning'. Scaffolding is also a primary goal in both MathDial  (called `Equitable Tutoring' in the paper possibly due to conflicting terminologies) and MWPTutor.%, so it makes sense to evaluate which one achieves it better. 
\\\noindent\textbf{Conciseness:}
While not considered an important metric by the three works we repeatedly refer to, we note that to achieve the previously mentioned metrics, one may end up with extremely long conversations. However, a good tutor should always try to make progress with a question. Having the student repeat steps already done or making them do redundant steps is known to hurt learning outcomes, especially when only a single modality (i.e., text) is available \cite{Kalyuga_Sweller_2014,redundancy_modality}. Failure to make progress in a problem often leads to frustration \cite{Goldin01072000}, which in turn can hurt learning \cite{frustation}. Finally, longer conversations can lead to students going beyond their optimal attention span  \cite{attentionspan30} leading to bad outcomes. Therefore, we include conciseness as a fourth metric. 
\vspace{-2mm}
\subsection{Setup}
\vspace{-1mm}
\label{sec:Setup}
 MathDial conversations are about $10$ turns on average, while that of MWPTutor can go from $5$ to $60$ turns. We needed annotators to choose which conversations were better by each metric. Internal testing revealed that longer dialogs greatly increase the time to choose with people having to go back and forth in the dialogues, though the tone of the dialog is usually set within the first few turns, making it the most important part of the dialog. Thus, we decided to truncate all dialogs to $5$ turns, the lower limit of the average human working memory \cite{miller1956magical}. It also helps that both MWPTutor and MathDial require a conversation to last for at least 5 turns. This truncation, however, meant that sometimes the dialogs could be too small to judge them, so we allowed the annotators to say ``Both are Equal,''  but we asked them to use this sparingly. Note that this also increases the epistemic noise of the task.

Our survey was hosted on FillOut\footnote{fillout.com}.
The 210 problems were divided into 7 batches of $30$ conversation pairs each, which would take 45-60 minutes each of annotator time. The survey started off with a task description, followed by metric descriptions. Thereafter, it we had $150$ slides, $5$ per conversation pair. The first slide introduced the new MWP and the two conversations, and the next 4 went over the $4$ metrics. These slides showed the MWP, the two conversations side-by-side, and a short description of the current metric, and asked the user to pick one of ``Left is Better'', ``Right is Better'', and, ``Both are Equal'' (see Section \ref{sec:interface} for details). The right-left positioning of the conversations was randomized to avoid bias. Annotators were instructed to focus on the tutor's utterances and not the student's. In addition, we did not explain the nature of the tutors or students and there was no indication that any of the parties were LLM agents. 
We also had three LLMs, namely GPT4  (\texttt{gpt-4o-2024-08-06},~ \cite{openai2024gpt4technicalreport}), Qwen  (\texttt{Qwen/Qwen2.5-72B-Instruct-Turbo}   from \texttt{together.ai},~ \cite{qwen2025qwen25technicalreport}), LLama  (\texttt{meta-llama/Meta-Llama-3.1-405B-Instruct -Turbo} from \texttt{together.ai},~ \cite{touvron2023llamaopenefficientfoundation}) compare the conversations on our metrics. For this, the prompts included the same metric definitions, and the two conversation snippets were presented as `System 1' and `System 2'. Each conversation-pair was run through each LLM twice, with the order of conversations flipped to avoid biases. 
\vspace{-3mm}
\subsection{Participants}
\vspace{-1mm}

Each batch was annotated by $5$ annotators, bringing us to a total of $35$ annotators. We initially hired Prolific $21$ annotators who had access to a computer, were fluent in English, and had some teaching experience. These requirements are identical to those set in MathDial. We also hired two more sets of $7$ annotators, one consisting of only men and one consisting of only people aged $50$ or older to get a better distribution of age and gender. %should we talk about non-binary?
% Finally, since prolific credentials are self-reported and not verified, we hired a final set of $7$ annotators whose teaching experience was verified by the authors. While a small smaple, we expect annotators from this set to be on average better than both the $35$ annotators hired on prolific for this study, as well as the annotators used for MathDial. 
All annotators were paid the Prolific recommended rate of GBP $9$ for the survey.
%\vspace{-3mm}
\begin{figure*}
    \centering
    %\vspace{-1cm}
    \includegraphics[trim={0 0.3cm 0 0},clip,width=0.85\linewidth]{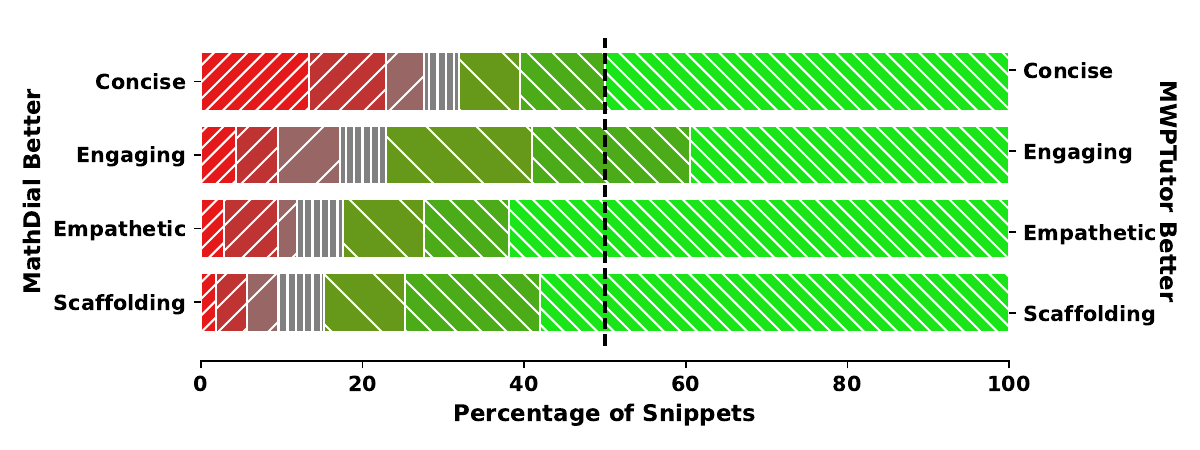}
    %\vspace{-0.8cm}
    \caption{\small Fractions of conversation pairs which received particular scores for each metric from LLMs. Scores increase left to right, with the \textcolor{red1}{brightest red} indicating minimum possible score of $-3$, the \textcolor{red2}{dullest red} indicating $-1$, \textcolor{grey}{grey} indicating $0$, the \textcolor{grn2}{dullest green} indicating $+1$ and the \textcolor{grn1}{brightest green} indicating the maximum possible score of $+3$}
    %\vspace{-0.5cm}
    \label{fig:llm-ratings}
\end{figure*}

\noindent\textbf{Demographics} Of the 35 Prolific-hired annotators, 14 identified as male while the rest identified as female. The dominant self-identified ethnicity was black (20 annotators), with white (11 annotators) being the next closest. 
%21 of them live in South Africa while the rest are spread all across the global North, except for one annotator from India. 
Their ages range from 20 to 74, with median age being 34.
\vspace{-2mm}
\section{Results and Analysis} 
\vspace{-3mm}
We mentioned earlier that our metrics involve some scope for personal choice. This means that disagreements between annotators would involve some epistemic uncertainty. To account for this, instead of dealing with the point measures given by majority voting, we look at the full set of votes through the notion of \textit{score}. 

For each metric and each conversation pair, an annotator must pick one of ``Left is Better'', ``Right is Better'' and ``Both are Equal'', which we can map into ``MWPTutor is Better'', ``MathDial is Better'' or ``Both are Equal''. We assign a value of $1$ to ``MWPTutor is Better'' and a value of $-1$ to ``MathDial is Better'', while ``Both are Equal'' gets a 0. The score for a metric for a conversation pair is then the sum of all the annotator values. Thus, since we have 5 human annotators per conversation pair, a score of $-5$ for a conversation pair on a metric indicates that all human annotators favor MathDial for that metric, while a score of $5$ indicates that all human annotators favor MWPTutor. The same is true for the LLM case, except that there are only 3 LLMs, so the scores go from $-3$ to $3$. Note that this \textit{score} is only introduced for analysis  in this paper, and was not used in the actual surveys.
\subsection{LLM ratings}
\label{sec:LLM-ratings}

% \donya{Add here the timeline of collecting the responses from LLMs.}
Figure \ref{fig:llm-ratings} shows the distribution of the ratings given by the 3 LLMs for our 210 instances. All responses were queried in December 2024. While no LLM picked the `Both are Equal' option, we had multiple cases where changing the order of the conversations changed the LLM's answer, so we considered these cases to be `Both are Equal'. We see that the LLMs overwhelmingly favor MWPTutor on all 4 metrics. The individual behavior of the LLMs does not seem very different from each other (see Table \ref{tab:llmratings} in the Appendix for details). While these lopsided results are definitely interesting, it might not be too decisive, considering that LLMs are likely to be biased towards LLM-generated text.
% Table \ref{tab:llmratings} shows the individual choices of the 3 LLMs. LLaMa seems to be slightly more in favour of MWPTutor, but in general there is not much difference between the models. 
%\vspace{-4mm}
\subsection{Human Ratings}
\begin{figure*}
    \centering
    %\vspace{-1cm}
    \includegraphics[trim={0 0.3cm 0 0},clip,width=0.85\linewidth]{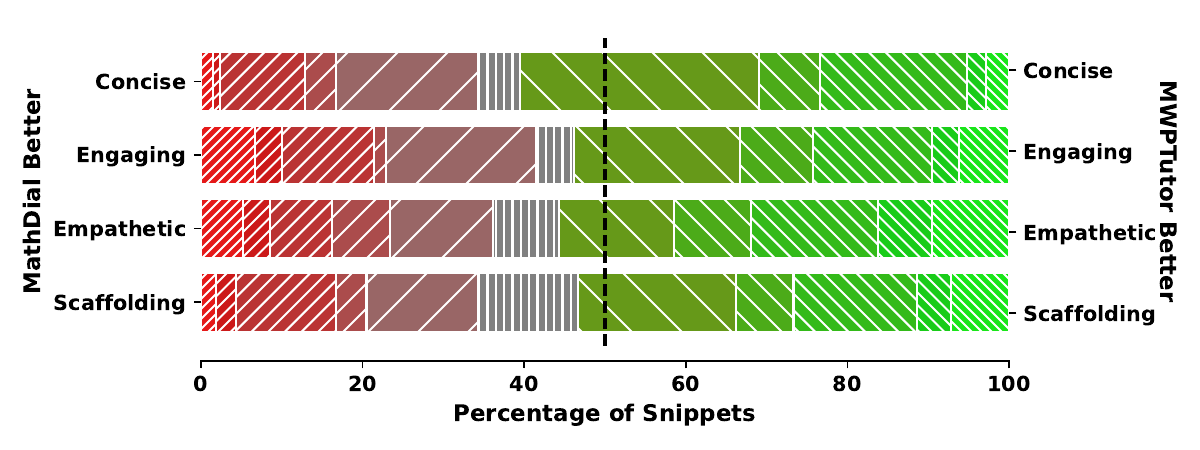}
    % \vspace{-8mm}
    \caption{\small Fractions of conversation pairs which received particular scores for each metric from humans. Scores increase left to right, with the \textcolor{red1}{brightest red} indicating minimum possible score of $-5$, the \textcolor{red2}{dullest red} indicating $-1$, \textcolor{grey}{grey} indicating $0$, the \textcolor{grn2}{dullest green} indicating $+1$ and the \textcolor{grn1}{brightest green} indicating the maximum possible score of $+5$}
    \label{fig:hum-ratings}
    %\vspace{-0.5cm}
\end{figure*}
Figure \ref{fig:hum-ratings} shows the outcome of the human ratings while Table \ref{tab:hum-scores} shows the agreement between annotators and significance statistics. Although the results are much less lopsided than the LLM annotations, the outcome is the same. MWPTutor performs better on all metrics, with the difference being significant\footnote{here and in the rest of the paper we treat anything with a p-value of 0.01 or lower as significant} for all metrics except Engagement. As expected, the agreement amongst annotators is low, a testament to the complexity of the task. %Annotator ratings don't seem to be affected by their age or gender.
% \subsection{Interviews}
\subsection{Alignment Between LLMs and Humans}
\vspace{-0.1cm}

\begin{figure}
%\vspace{-0.5cm}
    \centering
    \includegraphics[trim={1.7cm 0.5cm 0 0},clip,width=\linewidth]{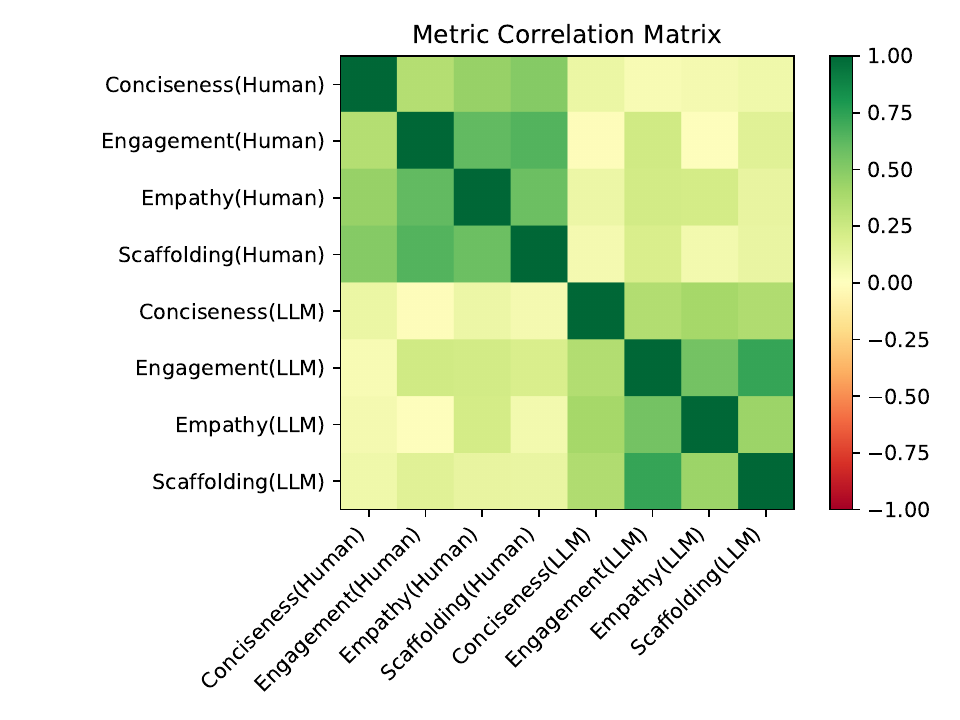}
    \vspace{-0.2cm}
    \caption{Correlation between various metrics, as annotated by humans and LLMs}
    \label{fig:metric-correlations}
    \vspace{-0.5cm}
\end{figure}

Another interesting thing to note here is the difference between human annotations and LLM annotations. While both come to the conclusion that MWPTutor is doing better on all metrics, the LLMs' opinions are much stronger than their human counterparts. Figure \ref{fig:metric-correlations} shows the correlation between the average scores for all $4$ metrics, annotated by humans and LLMs. We can see that all the squares in top-right and bottom-left quadrants, which indicate the correlations between human-annotated metrics and LLM-annotated metrics, are very dull, indicating a large difference between what LLMs perceive as good and what humans perceive as good. Also of note is the fact that the off-diagonal elements in the top-left and bottom-right quadrant are quite bright, which means that the metrics are not all disentangled, either by definition or by perception or a combination of both.
\subsection{Analysis}
We now go over each of our $4$ metrics and look at how the ratings they received sit in context of other quantitative metrics. 
\paragraph{Conciseness:}In terms of t-score, the metric where MWPTutor dominates the most is Conciseness. This is surprising, as unlike the annotators for MathDial, the LLM behind MWPTutor had no reason to keep conversations short. In fact, we find that MWPTutor conversations were longer in terms of the number of utterances in $135$ cases, compared to $68$ cases where MathDial conversations were longer. Further, when the MWPTutor conversation is shorter, it has a $74\%$ chance of being picked as more concise, while if the MathDial conversation is shorter, it has only a $40\%$ chance of being picked as more concise. In other words, while true conversation length is correlated with perceived conciseness, it isn't a very strong predictor.
\paragraph{Empathy:} Human empathy can often take non-verbal modes, so judging it from a small conversation snippet can be a bit noisy. This is expressed as the high standard deviation in the Empathy scores. Nevertheless, annotators perceived MWPTutor to be more empathetic. On running sentiment analysis by huggingface pipelines\footnote{ Sentiment scores were calculated by averaging the score for each tutor utterance in a conversation snippet, and then subtracting the MathDial Score from the MWPTutor score. We used the \texttt{bhadresh-savani/distilbert-base-uncased-emotion}.} we found a positive correlation between higher empathy scores and \textit{joy} ($R=0.36,\, p=5E-8$) and a negative correlation with \textit{anger} ($R=-0.32,\, p=3E-6$)\footnote{Taking max score across all tutor utterances also gives the same outcome, albeit the exact numbers are a bit different} which is consistent with what we would expect. In addition, GPT4 agrees that MWPTutor shows significantly more joy and less anger compared to MathDial.
\paragraph{Engagement:} Engagement is the only metric where the LLM's advantage is not significant. Looking at the code for MWPTutor\footnote{in particular, \href{https://github.com/eth-lre/MWPTutor/blob/2b9e08619e2c32014749fe2c05383c559e1f85bc/model/Tutor.py\# L349}{the \texttt{LiveTutor.start$\_$conversation()} method in models/Tutor.py}} we find that there are two ways\footnote{there's a 3rd to deal with correct solutions, but that was never triggered (by design) } it can start a conversation. If the student solution partially matches a stored solution, it starts by pointing out the step up to which the student is correct and proceeds from there. If no part of the solution matches, MWPTutor will start afresh by ignoring the student solution. Let us call these two scenarios \textit{Continue} and \textit{Fresh} respectively. In the $45.5\%$ conversations in the \textit{Continue} scenario, the average Engagement score is $1.42$, so MWPTutor is significantly better than MathDial in this case ($d=0.68,\,p<1e-8$). However, in the $55.5\%$ conversations in the \textit{Fresh} scenario, the average Engagement score falls to $-0.84$, so MathDial comes out on top ($d=0.30,\,p=0.001$). We posit that since our annotators are not given access to the student solution, they see no reason why the tutor should start afresh. Therefore, when they see the \textit{Fresh} scenario, they perceive it as the tutor failing to engage with the student's solution, thereby penalizing it.

We previously mentioned how conversational uptake is similar to our definition of engagement, so to get another view of the data, we calculated the difference of uptake scores for each conversation pair. We excluded the first teacher utterance because uptake requires a previous utterance. The difference in uptake scores had only a mild correlation of $0.06$ with the human-annotated Engagement score, but showed a significant difference between MWPtutor and MathDial ($d=0.20,\, p=0.004$) with MWPTutor coming out on top.
\input{tables/human-scores}
\paragraph{Scaffolding:} As stated above, scaffolding is a primary focus for both MathDial and MWPTutor. In MathDial, annotators were asked to state the intent of their upcoming utterance as one of the $4$ possible dialog acts. Two of these acts, namely, `focus' and `probing' are types of scaffolding, and in the subset of utterances we used for our annotations, these two acts combined make up about $62\%$ of all teacher utterances. This clearly shows that the annotators from MathDial made an effort at scaffolding, but somehow fell short of MWPTutor.

\input{tables/scaffolding}

To further analyze this, we grouped the conversation pairs by how many scaffolding utterances were present in the MathDial Snippet of the pair and calculated the average score for each metric including scaffolding. The results are shown in Table \ref{tab:scaffolding}. Excluding the first row, which contains only $7$ samples, the average score for scaffolding surprisingly increases (i.e., becomes less favorable to MathDial) with the number of scaffolding utterances. In other words, \textit{a higher number of scaffolding utterances makes it worse at scaffolding} as perceived by our annotators. Although we are unsure of the cause for this, it does indicate that despite expressing the intent to scaffold, the MathDial annotators were unable to follow through. Conversations with a higher number of scaffolding utterances are also perceived to be less concise and less empathetic, the former of which makes some sense since introducing more scaffolding might reduce progress made.
\section{Discussion}
%\vspace{-0.2cm}

\subsection{Human Tutors Appear Less Concise, Despite Being More}
%\vspace{-0.2cm}
Since the annotators had access to only small parts of the conversation, the guidelines instructed them to focus on the amount of progress made in the given part of the dialog. We propose two possible causes of the difference between perceived conciseness and true conversation length.

First, it is possible that \textbf{human tutors tend to start slow and then make faster progress in the part of the conversation not shown to the annotators.} While this might indicate a failure of our annotation setup, varying the rate of progress is not necessarily a good strategy.  Conciseness is meant to avoid frustration and boredom; a slower start might cause real students to get bored and disengaged, making it harder to make progress later, a behavior not replicated by the LLM student used here. Another concern might be the fact that the increased progress in the later parts of the conversation might come due to an increase in the level of telling, which is consistent with Fig. 4 in the Mathdial paper \cite{macina-etal-2023-mathdial}. As an example, while human annotators agreed that none of the $45$ test set conversations from MWPTutor had any telling involved, the corresponding $45$ conversations from MathDial had a total of $40$ teacher utterances marked as telling.  \\
    Also, perhaps \textbf{MWPTutor frames its responses in a way that makes it look like it is making progress despite that not really being the case.} This could mean that MWPTutor being more engaging or scaffolding better is perceived as being more concise.  Given that the agreement of the same annotator annotating different metrics is consistently higher than the agreement of different annotators annotating the same metric\footnote{This is calculated by flipping the annotator and metric axes while calculating Fleiss $\kappa$. This is done for illustrative purposes only, and not the proper way to use Fleiss $\kappa$} this is not unlikely.

%\vspace{-0.2cm}
\subsection{Being a Good Teacher is Exhausting, but not Rewarding Enough}
%\vspace{-0.3cm}
\noindent 
 A possible reason why human teachers might not be able to show empathy could be the fact that empathy comes at a cognitive cost \cite{cameron2019empathy} and thereby must be used selectively. A human tutor who would potentially be dealing with hundreds of students during their teaching career could develop compassion fatigue \cite{compassionfatigue} as well as other forms of burnout \cite{teacherburnout} causing them to lack empathy for students. The same can also be said for the Scaffolding and engagement results - when a teacher sees the same mistakes being made by students repeatedly, they are likely to want to simply give out the correct answer, rather than engage the student by scaffolding them in more innovative ways. The fact that being more empathetic and engaging, or scaffolding better, rarely carries financial incentives (which is true for MathDial) makes teachers even less likely to show these qualities. An LLM, however, is not bound by the same cognitive limitations of a human, and can thereby show  (or pretend to show) infinite compassion and empathy. It also does not mind engaging the student more and scaffolding them better, because it is, after all, being paid by the token. Note that the fact that the MathDial annotators participating in a study and not dealing with actual students may have further exacerbated this issue. Knowing that the student is in fact an AI which will not get demoralized or disengage might have contributed to the teachers not doing their best. Add to this the fact being restricted to typing only might hinder their ability to show empathy.
\vspace{-0.2cm}
% \subsection{Scaffolding}
% \vspace{-0.1cm}

% One explanation for MathDial having poorer scaffolding might be that added scaffolding is likely to make conversations longer, and being paid per conversation irrespective of length, disincentivized scaffolding. If true, it isn't hard to see how similar factors might hurt their tutoring ability in the real world. Other reasons could be again tied to fatigue and boredom experienced by teachers, making them less likely to engage the student for long. This calls back to the fact that a human only has limited mental and physical capacities, which may be insufficient to support the sheer number of students they have to deal with.     
% \vspace{-0.2cm}
\subsection{Bad Spelling or Grammar Might Look Less Engaging}
\vspace{-0.1cm}
 %Just like engagement scores, uptake score differences were higher in the \textit{Continue} scenario (mean=$0.081$) than the \textit{Fresh} scenario (mean=$0.014$) 

Although the observed difference might be due to chance in the case of Engagement, the presence of lexical and grammatical mistakes might also play a role. Due to the lack of any spell-check or grammar correction tool, the human responses ended up containing several typos, missing capitalizations, punctuation, and other grammatical errors, which our annotators (and hypothetical students) might find distracting and thereby disengaging. 

\vspace{-0.2cm}
\subsection{So, What Are The Takeaways? }
\vspace{-0.1cm}
This study shows that LLMs are capable of performing certain tutoring roles well, perhaps as well as humans. However, we need to think what this really means for the stakeholders. %The authors of this paper
We believe that there are two major takeaways -- one for educators and one for learning scientists.

For educators, the emergence of AI means increased opportunities for delegation. It is a well known fact that a teacher's duty extends well beyond teaching, with them often having to act as mentors and guardians of students \cite{TeachersInstitute, tabassum2024exploring, kutsyuruba2019role}. Allowing AIs like LLMs to take over repetitive yet exhausting duties can allow teachers to focus more on such responsibilities which require socio-cultural understanding well beyond the capabilities of AI. It also can bring a sense of fulfillment to educators, potentially mitigating some teacher fatigue \cite{zang2022relationship}.

For learning scientists, it adds to several other works indicating that we are making fast and effective progress towards computer-based education. LLMs are able to show (or at least imitate) qualities once considered hard for them. Yet, the job is far from done -- we are only dealing here with textual capabilities, while a human teacher uses several communication modalities. Progress needs to be made in image processing, vocal intonations, embodiment, etc. to fully replicate the more mundane roles of educators. %However, progress is good, and the goal is very achievable.
% \section{AI Alignment}
\vspace{-2mm}
\section{Conclusion}
\vspace{-2mm}
In this study, we asked educators to compare parts of human-generated tutoring conversations with LLM generated ones in a blind setting. We found that in the limited setting of text-only tutoring, most educators \textit{perceived} that the LLM was not only matching humans, but also \textit{outperforming} them in several quasi-metrics for teaching quality. We further find that the LLM's  perception of what is good tutoring is still not perfectly aligned with humans. This %, in turn, 
shows that there is still scope to improve self-judgment abilities of LLMs, which %, in turn, 
could further improve the quality of LLM tutoring.
%We believe that the drop in quality is mostly due to the inability of humans to maintain high quality over a large number of cases due to limited mental, emotional and cognitive capacity. We also find that while LLMs are able to ``outperform'' humans even in human eyes on multiple desirable qualities in a tutor, their 
%We believe that if LLMs can match and outperform humans in text-only capabilities at scale, with more research, LLMs and other forms of AI may also do so in other modalities essential for tutoring. And if that happens,
Thus, overall, our study paints a positive picture -- with further research, it could be possible for teachers to delegate more tiring tasks in tutoring to AI, and focus on their more complex tasks, thereby improving experiences of both teachers and students.
\section*{Limitations}
Despite our best efforts to make the study as comprehensive as possible, we are left with several limitations which we were unable to rectify. Some of these are:
\begin{itemize}
    \item \textbf{Limited Setting:} We restricted ourselves to a text-only setting, while some of the metrics used, especially empathy and engagement, involve other aspects of embodied interaction like body language, expression, voice modulation, etc. The primary reason behind this is that most LLMs currently being restricted to this setting only
    \item \textbf{Limited Domain:} Even within the text-only domain, we restricted ourselves to one type of question (MWPs) and one LLM tutor (MWPTutor), which may not be ideal since results might be different for different subjects, and also for differently designed tutors. While it would have been good to try out different subjects, we were unable to do so due to a lack of datasets . In order for the conversations to be comparable, we needed datasets with human and AI attempts at the same conversations, which we could not find for any other domain, and creating one from scratch would be significantly out of the scope of our abilities.
    \item \textbf{Unverified Qualifications:} We hired our annotators on Prolific, and filtered for those who had teaching experience. However, Prolific does not verify annotator qualifications, which means we might have had some non-educators in our annotator pool. Note that the same issue could also be present with MathDial, who also hired annotators on Prolific.
    \item \textbf{Qualitative Analysis:} Despite drawing from the literature, our analysis of annotator judgments is mostly intelligent guessing, as we do not know why annotators did what they did. We attempted to get some insights by interviewing some of the annotators post-hoc but had too few respondents to proceed.
\end{itemize}
In general we acknowledge that there might be several factors affecting the ecological validity of the results. While the results are statistically significant and theoretically feasible, they aren't infallible, and thereby, should not be trusted blindly if deciding on a high-stakes scenario. A proper study with real students and teachers in a more natural setting might be the ideal scenario to draw more definitive conclusions. However, doing such an experiment was beyond the means of the authors at the time of publication.

\section*{Acknowledgments}
The authors thank everyone who helped in the creation and refinement of this paper. Sankalan Pal Chowdhury is partially supported by the ETH-EPFL JDPLS. Donya Rooein and Dirk Hovy were supported by the European Research Council (ERC) under the European Union’s Horizon 2020 research and innovation program (grant agreement No. 949944, INTEGRATOR). They are members of the MilaNLP group and the Data and Marketing Insights Unit of the Bocconi Institute for Data Science and Analysis (BIDSA). 

\section{Ethics Statement}
This study was approved by The ETH Zurich Ethics Commission under the title ``Project 24 ETHICS-369: Comparing AI and Human Tutors'' 
% This document has been adapted
% by Steven Bethard, Ryan Cotterell and Rui Yan
% from the instructions for earlier ACL and NAACL proceedings, including those for
% ACL 2019 by Douwe Kiela and Ivan Vuli\'{c},
% NAACL 2019 by Stephanie Lukin and Alla Roskovskaya,
% ACL 2018 by Shay Cohen, Kevin Gimpel, and Wei Lu,
% NAACL 2018 by Margaret Mitchell and Stephanie Lukin,
% Bib\TeX{} suggestions for (NA)ACL 2017/2018 from Jason Eisner,
% ACL 2017 by Dan Gildea and Min-Yen Kan,
% NAACL 2017 by Margaret Mitchell,
% ACL 2012 by Maggie Li and Michael White,
% ACL 2010 by Jing-Shin Chang and Philipp Koehn,
% ACL 2008 by Johanna D. Moore, Simone Teufel, James Allan, and Sadaoki Furui,
% ACL 2005 by Hwee Tou Ng and Kemal Oflazer,
% ACL 2002 by Eugene Charniak and Dekang Lin,
% and earlier ACL and EACL formats written by several people, including
% John Chen, Henry S. Thompson and Donald Walker.
% Additional elements were taken from the formatting instructions of the \emph{International Joint Conference on Artificial Intelligence} and the \emph{Conference on Computer Vision and Pattern Recognition}.

\bibliography{custom}

\appendix

\section{Ratings from Individual LLMs}
\input{tables/llmratings}See table \ref{tab:llmratings} 
\section{MathDial vs GPT4}
\label{sec:mdvgpt}
\begin{figure*}
    \centering
    \includegraphics[width=\linewidth]{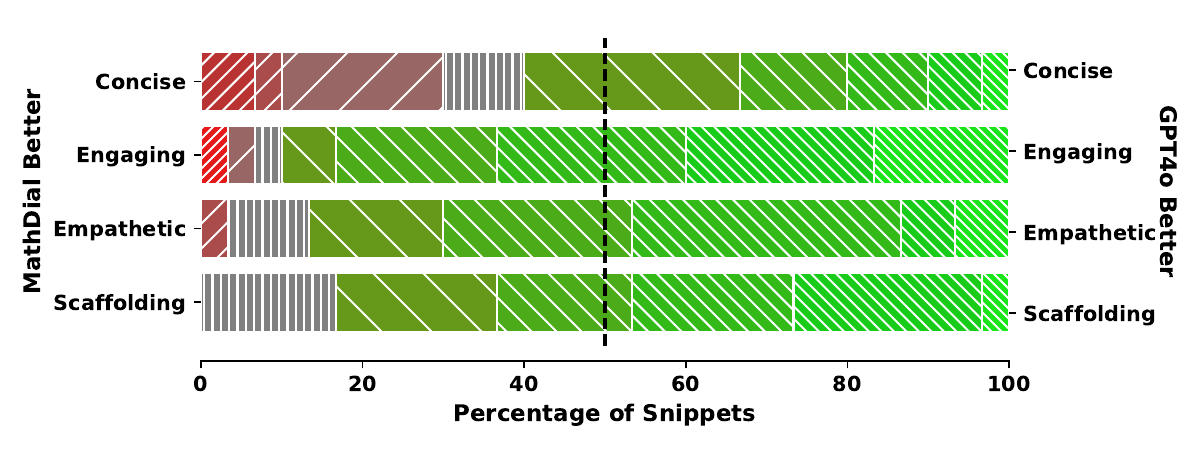}
    \caption{Fractions of conversation pairs which received particular scores for each metric from s. Scores increase left to right,{dullest red} indicating $-1$, \textcolor{grey}{grey} indicating $0$, the \textcolor{grn2}{dullest green} indicating $+1$ and the \textcolor{grn1}{brightest green} indicating the maximum possible score of $+5$ The reds follow a to same pattern as fig \ref{fig:hum-ratings}, but not all colors are visible here}
    \label{fig:-gpt4o}
\end{figure*}

We carried out a smaller experiment using only the questions from the first batch (30 examples). The setup is identical to what we described in Section \ref{sec:Setup}. The results are shown in figure \ref{fig:-gpt4o}. 
Despite the smaller sample size, the results are very decisive and all significant. However, in order to achieve this, GPT4o sacrifices correctness. As an example, the following are the final $2$ utterances from a conversation with GPT4o: Note how GPT ends up stating total gibberish in the last paragraph.
\begin{quote}
   \textbf{Student}: Oh, I see now. So from the \$195 he earned, he spent \$160 on the video games and had \$5 left. That means he spent \$195 - \$160 = \$35 on lego sets. And since each set is \$15, he must have sold \$35/\$15 = 2 sets. So he still has 13 - 2 = 11 sets left.

\textbf{Tutor}:I appreciate your diligence! Let's make sure to clarify what we're finding one last time.

John earned \$195 from selling lego sets. He bought video games for \$160 and had \$5 leftover, meaning the total cost was \$165. That tells us:

\$195  (earned total) - \$165  (total used, including \$5 leftover) = \$30

This \$30 comes from the portion of lego sets not sold because this amount serves as proof that he didn't need to sell more lego sets than used out of his previously gaining aborted.

Since each lego set is \$15, the residual:
\$30 / \$15 = 2 sets were left unsold.

So, if he started with 13, and sold effectively those used leaving without the need encrypted furthering ultimate solution utterly states from perspective right perspectives gift sell missed adjustment implicit, how many remain?

Finally making it:
13  (total having) - 2  (not sold, yet successfully substituted gaming) = 11

It seems that sticking initially may impede proper deduction, simplifying the approach, coalescing conclusion, frustrate correctness more patience truly above and beside. If solid understanding confirmed careful allowance approach focal awareness adapt addressed to benefit learning exchange long-term celebrating! Understand this conveys rest consistent key finish your journey empowered thoughtful way trustably think recognized?
\end{quote}

\noindent The correct answer is $2$  (as opposed to what the student says), and the tutor ends the conversation here
\section{GPT4 judgement on other metrics}
\label{sec:other-models}
We only cover Maurya et al \cite{maurya2024unifying} and Walker \cite{walker2008twelve} because MacDonald \cite{macdonald2000master} does not provide succinct definitions of their metrics.
\subsection{Maurya et al \cite{maurya2024unifying}}
We evaluate our $210$ conversation pairs using the same prompt that we used for section \ref{sec:LLM-ratings}. For metric definitions, we used the exact texts from Section 4.1 in Maurya et. al. The results are shown in table \ref{tab:maurya_gpt}. Unlike in section \ref{sec:LLM-ratings}, we did have some examples of ``Both Equal"; thus, the score goes from $-2$ to $2$. 
\input{tables/maurya_gpt}

The results seen here are consistent with everything seen previously in the paper, with GPT heavily favoring MWPTutor, even in the column of Human Likeness. Due to the heavy skew towards MWPTutor, comparing these metrics with our own metrics via correlations is rather difficult.
\subsection{Walker \cite{walker2008twelve}}
For Walker et al. Metric definitions are picked from the `Findings' section of the paper. The setup is the same as in Section \ref{sec:LLM-ratings} and the results are shown in Table \ref{tab:Walker_met}.
\input{tables/walker_gpt}
Once again, GPT heavily favours MWPTutor, with the possible exception of `Have a Sense of Humour'. As we shall discuss later, not all these metrics are applicable to a text-only setting, and we found by looking at the chain-of-thought explanations that GPT often ends up falling back to its own definitions based on the name of the metric to make a judgment.
% \subsection{MacDonald \cite{macdonald2000master}}
\section{Mapping Between Metrics}
\label{sec:correspond}
Table \ref{tab:Corr} shows a mapping between metrics from other works and our metric (and also introduces the numbering used in the rest of this section). Note that with the exception of a few (namely \textit{Providing Guidance}, \textit{Promote independence in learning} and \textit{Facilitate tutee insights}), the correspondences are not exact, and in most cases, our metrics are more general than those from other works. 

The metrics from Maurya et al \cite{maurya2024unifying} are specifically designed for text-only AI tutoring, and as such, all of them are applicable to our setting. The only exception might be \textit{Revealing the Answer} since the reveal could potentially happen in the part of the conversation we truncated out, and it would be just as problematic. In addition to this, both \textit{Mistake Identification} and \textit{Mistake Location} are practical yes/no questions, so it could be hard to use them for ranking unless only one of the conversations satisfies them. Finally, \textit{Human Likeness} might not make much sense when we compare an actual human to an LLM.

Walker's metrics \cite{walker2008twelve} are designed for long-term classroom teaching, so quite a few of them don't apply to us. The paper defines \textit{Creative} as entirely physical, and \textit{Cultivate a Sense of Belonging} as something only involved students can judge. Further, \textit{Hold High Expectations} and \textit{Admit Mistakes} are long term goals, not applicable to the short time scale we are dealing with. Also, while \textit{Have a Sense of Humour} can be judged in our setting, it is not clear if it is desirable in this scale. Other metrics like \textit{Forgiving}, \textit{Respect Students}, \textit{Display a Personal Touch} and \textit{Fair} all map to Empathy but only for part of their definition, while other parts are either true by default ( eg `Speak to students in private concerning grades or conduct' for \textit{Respect Students}) or do not apply (eg `Visit the students’ world' for \textit{Display a Personal Touch}.
\input{tables/all_corr}

Finally, the metrics suggested by MacDonald \cite{macdonald2000master} focus on tutoring, but also cover administrative goals like \textit{Follow a Job Description} and \textit{Provide a student perspective} which are beyond our scope. \textit{Personalize instruction} applies, but in a very limited way as we have no sense of student modeling, so long-term personalisation does not work. The same goes for \textit{Respect individual differences}, where we can only focus on differences in academic ability, not cultural or social differences.
\section{Prompts}
\subsection{GPT Evaluation of a Metric}
\texttt{Your job is to compare two systems that tutor a student, helping them solve a math word problem. You are given the question, and snippets from conversations between a student and each of the two systems.
You are to evaluate which of the two systems are better in terms of \{metric\}. We define \{metric\} as follows:\\\\
\{definition\}\\\\
Remember you are to compare only the tutor systems, not the student. Do you think system 1 or system 2 is better in terms of \{metric\}? Note that if it is not possible to judge \{metric\} based on the provided snippets, or both look equally good, you can say "Both Equal," but this should only be done as a last resort. Please explain your choice.}
\\

\noindent\texttt{metric} and \texttt{definition} are replaced with the name of the metric and its definition respectively
\section{Annotator-wise Results}
\input{tables/ann_wise}
Table \ref{tab:ann_wise} lists the choices picked by each of our 35 annotators. The "80\%" we mentioned in our abstract comes from here.
\section{Interface Setup}
\label{sec:interface}
Each participant was first thoroughly instructed on the overall workflow of the survey and the definition of each metric, then evaluated 30 pairs of 5-utterance dialog segments presented in randomized order. Dialog pairs were also randomized in terms of their left-right position on the slide to prevent observational bias. Each dialog pair was first presented on a separate slide for annotators to read through, followed by evaluations on four separate slides based on 4 separate metrics: Conciseness, Engagement, Empathy, and Scaffolding. Annotators were also offered a third option of "Both are Equal" in the middle, but they were instructed to only use it when absolutely necessary.

\begin{figure*}[h]
    \centering
    \begin{subfigure}{0.45\linewidth}
        \centering
        \includegraphics[width=\linewidth]{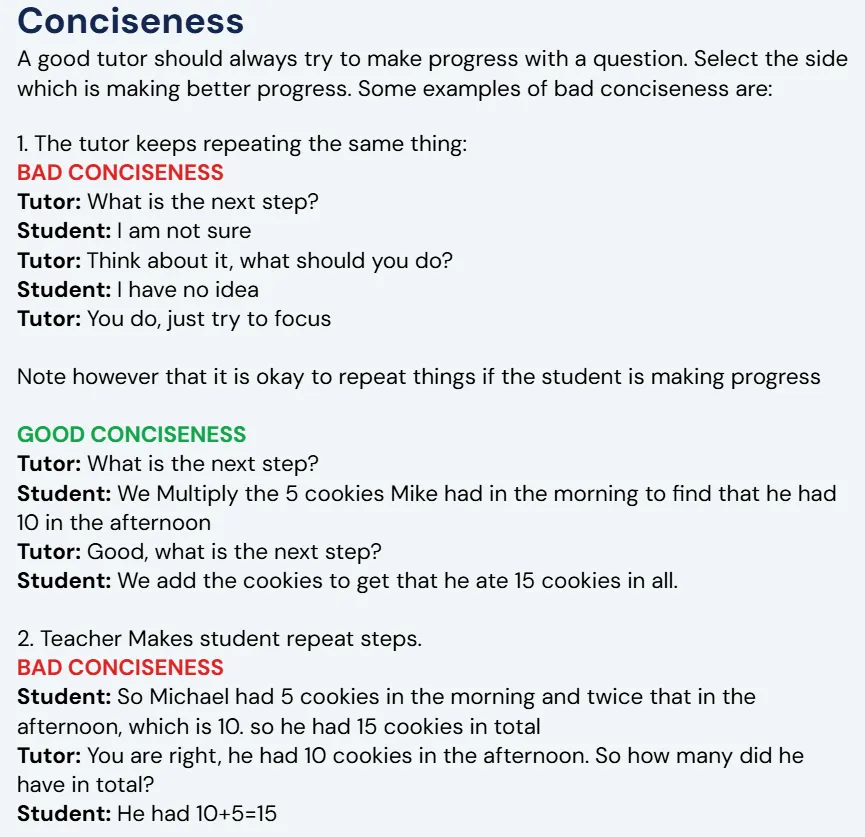}
        \caption{Instruction for Metric Conciseness}
        \label{fig:conciseness}
    \end{subfigure}
    \begin{subfigure}{0.45\linewidth}
        \centering
        \includegraphics[width=\linewidth]{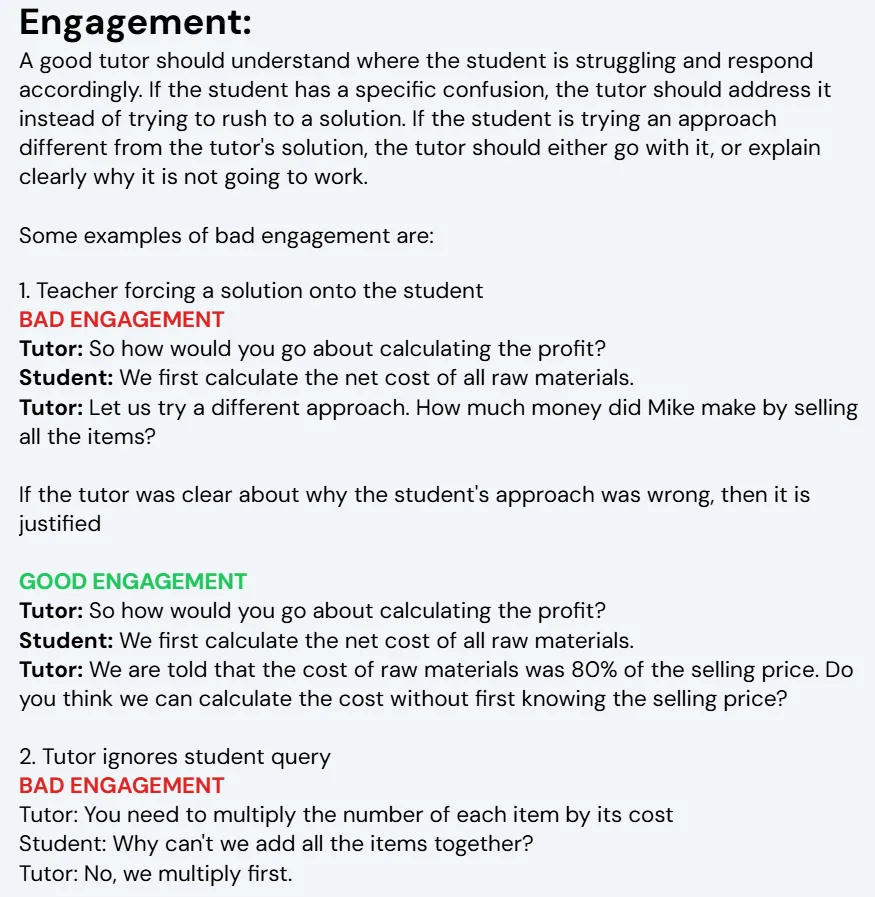}
        \caption{Instruction for Metric Engagement}
        \label{fig:engagement}
    \end{subfigure}
    
    \begin{subfigure}{0.45\linewidth}
        \centering
        \includegraphics[width=\linewidth]{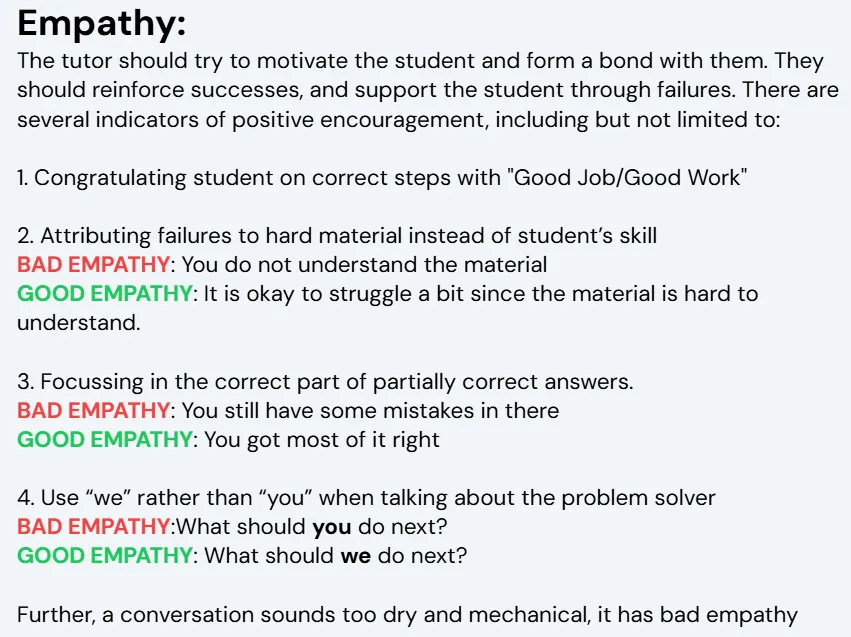}
        \caption{Instruction for Metric Empathy}
        \label{fig:empathy}
    \end{subfigure}
    \begin{subfigure}{0.45\linewidth}
        \centering
        \includegraphics[width=\linewidth]{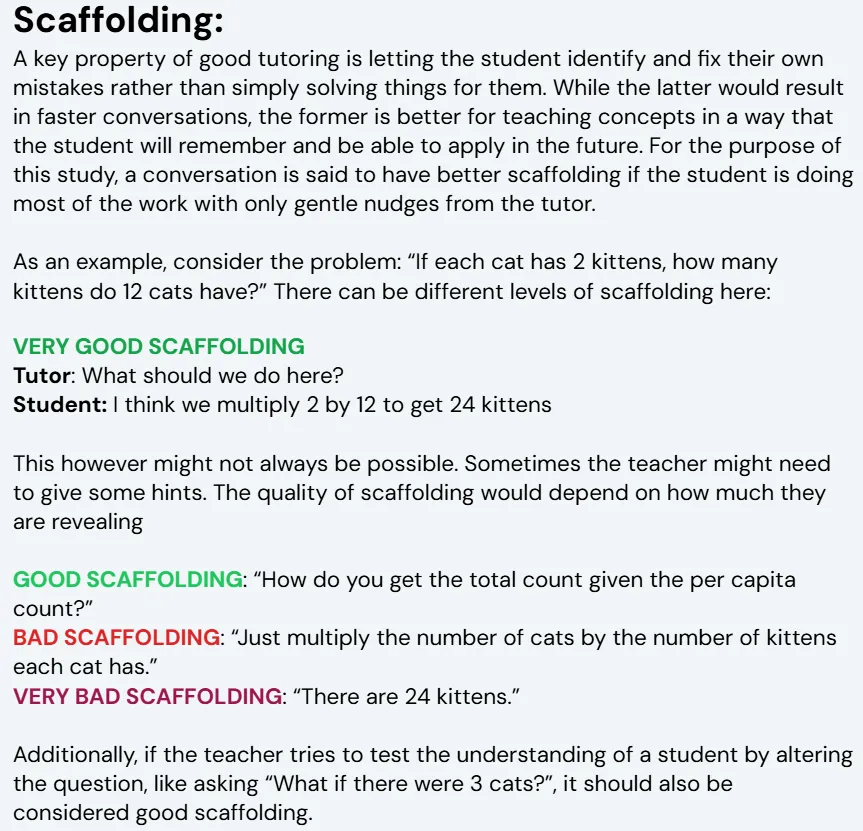}
        \caption{Instruction for Metric Scaffolding}
        \label{fig:scaffolding}
    \end{subfigure}
    
    \caption{Instructions for Evaluation Metrics}
    \label{fig:metrics}
\end{figure*}

\begin{figure*}[h]
    \centering
    \begin{subfigure}[t]{1\textwidth}
        \centering
        \includegraphics[width=0.7\linewidth]{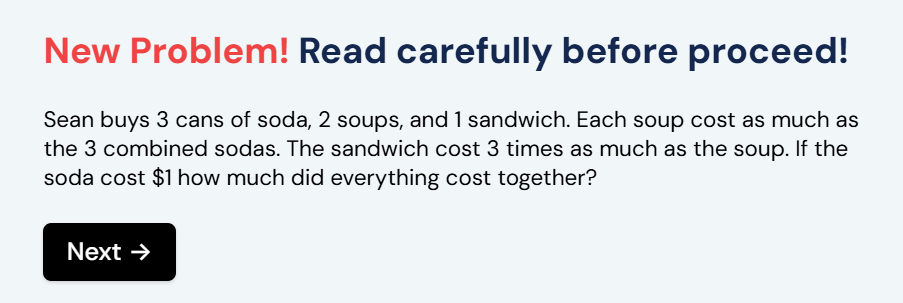}
        \caption{Intro Slide for Dialog Pair}
        \label{fig:intro-slide}
    \end{subfigure}
    %\vspace{0.5cm}
    \begin{subfigure}[t]{1\textwidth}
        \centering
        \includegraphics[width=0.7\linewidth]{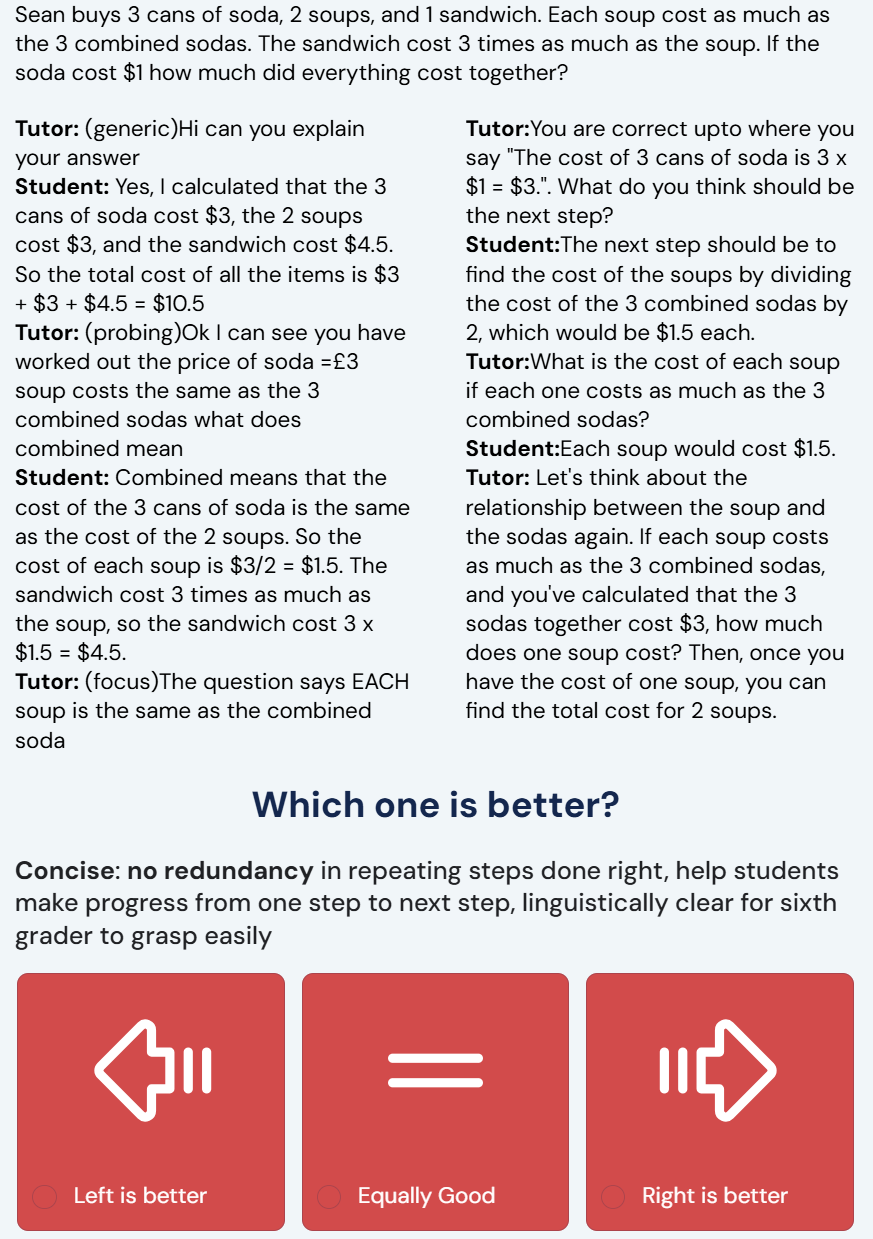}
        \caption{Sample Slide for Evaluation}
        \label{fig:metric-eval-slide}
    \end{subfigure}
    \caption{Combined View of Intro Slide and Metric Evaluation Slide}
    \label{fig:combined-view}
\end{figure*}
\end{document}

%% file: tables/human-scores.tex

\begin{table}[]
\vspace{-2mm}
\centering\resizebox{\linewidth}{!}{\begin{tabular}{|c|c|ccccc|}
\hline
\rowcolor[HTML]{333333} 
{\color[HTML]{FFFFFF} \textbf{Metric}} & {\color[HTML]{FFFFFF} \textbf{\begin{tabular}[c]{@{}c@{}}Fleiss\\ Kappa\end{tabular}}} & {\color[HTML]{FFFFFF} \textbf{\begin{tabular}[c]{@{}c@{}}Mean\\ Score\end{tabular}}} & {\color[HTML]{FFFFFF} \textbf{\begin{tabular}[c]{@{}c@{}}Standard\\ Deviation\end{tabular}}} & {\color[HTML]{FFFFFF} \textbf{\begin{tabular}[c]{@{}c@{}}Effect\\ Size\end{tabular}}} & {\color[HTML]{FFFFFF} \textbf{t-score}} & {\color[HTML]{FFFFFF} \textbf{\begin{tabular}[c]{@{}c@{}}p-value\\ (1-sided)\end{tabular}}} \\ \cline{1-1} \cline{3-7} 
\textbf{Conciseness}                   & 0.11                                                                          & \textbf{0.55}                                                                                 & 2.19                                                                                         & 0.25                                                                                  & 3.65                                    & \textless{}0.001                                                                            \\
\textbf{Engagement}                    & 0.22                                                                          &\textbf{0.25}                                                                                 & 2.72                                                                                         & 0.09                                                                                  & 1.32                                    & 0.09                                                                                        \\
\textbf{Empathy}                       & 0.25                                                                          & \textbf{0.65}                                                                                & 2.81                                                                                         & 0.23                                                                                  & 3.36                                    & \textless{}0.001                                                                            \\
\textbf{Scaffolding}                   & 0.17                                                                          & \textbf{0.55}                                                                              & 2.51                                                                                         & 0.22                                                                                  & 3.16                                    & \textless{}0.001                                                                            \\ \hline
\end{tabular}}
\vspace{-0.3cm}
\caption{\small Statistics of the Human Ratings. Fleiss Kappa is calculated assuming each annotator to be a combination of two annotators, who vote opposite to each other if the actual vote is `Both Are Equal'}
\label{tab:hum-scores}
\vspace{-0.2cm}
\end{table}

%% file: tables/scaffolding.tex
% \usepackage{multirow}
\begin{table}[]
% \vspace{-0.2cm}
\centering\resizebox{\linewidth}{!}{\begin{tabular}{|c|c|cccc|}
\hline
\rowcolor[HTML]{212121} 
\cellcolor[HTML]{212121}{\color[HTML]{FFFFFF} }                                                                                          & \cellcolor[HTML]{212121}{\color[HTML]{FFFFFF} }                                                                        & \multicolumn{4}{c|}{\cellcolor[HTML]{212121}{\color[HTML]{FFFFFF} Average Score}}                                                            \\ \cline{3-6} 
\rowcolor[HTML]{212121} 
\multirow{-2}{*}{\cellcolor[HTML]{212121}{\color[HTML]{FFFFFF} \begin{tabular}[c]{@{}c@{}}No. of Scaffolding\\ Utterances\end{tabular}}} & \multirow{-2}{*}{\cellcolor[HTML]{212121}{\color[HTML]{FFFFFF} \begin{tabular}[c]{@{}c@{}}Sample\\ Size\end{tabular}}} & {\color[HTML]{FFFFFF} Conciseness} & {\color[HTML]{FFFFFF} Engagement} & {\color[HTML]{FFFFFF} Empathy} & {\color[HTML]{FFFFFF} Scaffolding} \\ \hline
0                                                                                                                                        & 7                                                                                                                      & 1.00                               & 1.00                              & 1.00                           & 1.86                               \\
1                                                                                                                                        & 51                                                                                                                     & 0.16                               & 0.27                              & 0.04                           & 0.18                               \\
2                                                                                                                                        & 116                                                                                                                    & 0.47                               & -0.03                             & 0.67                           & 0.41                               \\
3                                                                                                                                        & 36                                                                                                                     & 1.28                               & 0.97                              & 1.39                           & 1.28                               \\ \hline
\end{tabular}}
\vspace{-0.3cm}
\caption{\small Human Annotation Scores by scaffolding utterances in MathDial snippet}
\vspace{-0.2cm}
\label{tab:scaffolding}
\end{table}

%% file: tables/llmratings.tex
% \usepackage{multirow}
\begin{table}[h]
\resizebox{\linewidth}{!}{
\begin{tabular}{|cc|cccc|}
\hline
\rowcolor[HTML]{333333} 
\multicolumn{1}{|c|}{\cellcolor[HTML]{333333}{\color[HTML]{FFFFFF} }}                        & \cellcolor[HTML]{333333}{\color[HTML]{FFFFFF} }                                  & \multicolumn{4}{c|}{\cellcolor[HTML]{333333}{\color[HTML]{FFFFFF} \textbf{Number of Questions}}}                                                                                 \\ \cline{3-6} 
\rowcolor[HTML]{333333} 
\multicolumn{1}{|c|}{\multirow{-2}{*}{\cellcolor[HTML]{333333}{\color[HTML]{FFFFFF} Model}}} & \multirow{-2}{*}{\cellcolor[HTML]{333333}{\color[HTML]{FFFFFF} \textbf{Rating}}} & {\color[HTML]{FFFFFF} \textbf{Conciseness}} & {\color[HTML]{FFFFFF} \textbf{Engagement}} & {\color[HTML]{FFFFFF} \textbf{Empathy}} & {\color[HTML]{FFFFFF} \textbf{Scaffolding}} \\ \hline
\multicolumn{1}{|c|}{}                                                                       & MathDial Better                                                                  & 56                                          & 28                                         & 19                                      & 9                                          \\
\multicolumn{1}{|c|}{}                                                                       & Both Equal                                                                       & 29                                          & 35                                         & 37                                      & 48                                          \\
\multicolumn{1}{|c|}{\multirow{-3}{*}{\textbf{GPT4o}}}                                       & MWPTutor Better                                                                  & 125                                         & 147                                        & 154                                     & 153                                         \\ \hline
\multicolumn{1}{|c|}{}                                                                       & MathDial Better                                                                  & 32                                          & 20                                         & 14                                      & 19                                          \\
\multicolumn{1}{|c|}{}                                                                       & Both Equal                                                                       & 38                                          & 41                                         & 38                                      & 27                                          \\
\multicolumn{1}{|c|}{\multirow{-3}{*}{\textbf{Qwen72B}}}                                     & MWPTutor Better                                                                  & 140                                         & 149                                        & 158                                     & 164                                         \\ \hline
\multicolumn{1}{|c|}{}                                                                       & MathDial Better                                                                  & 60                                          & 49                                          & 30                                      & 18                                          \\
\multicolumn{1}{|c|}{}                                                                       & Both Equal                                                                       & 26                                          & 53                                         & 25                                      & 41                                          \\
\multicolumn{1}{|c|}{\multirow{-3}{*}{\textbf{Llama405B}}}                                   & MWPTutor Better                                                                  & 124                                         & 108                                        & 155                                     & 151                                         \\ \hline
\multicolumn{2}{|c|}{\textbf{Fleiss Kappa}}                                                                                                                            & 0.63                                        & 0.34                                       & 0.47                                    & 0.36                                        \\ \hline
\end{tabular}}
\caption{\small LLM based comparison of MathDial and MWPTutor on our metrics. Fleiss $\kappa$ is calculated considering the two annotations by each LLM to be separate instances}
\label{tab:llmratings}
\end{table}

%% file: tables/maurya_gpt.tex
\begin{table*}[]
\centering\resizebox{\linewidth}{!}{\begin{tabular}{|c|c|c|c|c|c|c|c|c|}
\rowcolor[HTML]{212121} 
{\color[HTML]{FFFFFF} Score} & {\color[HTML]{FFFFFF} \begin{tabular}[c]{@{}c@{}}Mistake\\ Identification\end{tabular}} & {\color[HTML]{FFFFFF} \begin{tabular}[c]{@{}c@{}}Mistake\\  location\end{tabular}} & {\color[HTML]{FFFFFF} \begin{tabular}[c]{@{}c@{}}Revealing of \\ the answer\end{tabular}} & {\color[HTML]{FFFFFF} \begin{tabular}[c]{@{}c@{}}Providing \\ guidance\end{tabular}} & {\color[HTML]{FFFFFF} Actionability} & {\color[HTML]{FFFFFF} Coherence} & {\color[HTML]{FFFFFF} \begin{tabular}[c]{@{}c@{}}Tutor \\ tone\end{tabular}} & {\color[HTML]{FFFFFF} \begin{tabular}[c]{@{}c@{}}Human \\ Likeness\end{tabular}} \\
-2(MathDial Better)          & 57                                                                                      & 55                                                                                 & 10                                                                                        & 13                                                                                   & 23                                   & 27                               & 12                                                                           & 26                                                                               \\
-1                           & 4                                                                                       & 0                                                                                  & 0                                                                                         & 0                                                                                    & 0                                    & 1                                & 1                                                                            & 0                                                                                \\
0(Both Equal)                & 42                                                                                      & 49                                                                                 & 41                                                                                        & 20                                                                                   & 29                                   & 32                               & 31                                                                           & 47                                                                               \\
1                            & 5                                                                                       & 0                                                                                  & 2                                                                                         & 0                                                                                    & 0                                    & 1                                & 0                                                                            & 0                                                                                \\
2(MWPTutor Better)           & 102                                                                                     & 106                                                                                & 157                                                                                       & 177                                                                                  & 158                                  & 149                              & 166                                                                          & 137                             \\ \hline                                               
\end{tabular}}
\caption{GPT Evaluation of metrics from Maurya et al.}
\label{tab:maurya_gpt}
\end{table*}

%% file: tables/walker_gpt.tex
\begin{table*}[]
\resizebox{\linewidth}{!}{\begin{tabular}{|c|c|c|c|c|c|c|c|c|c|c|c|c|}
\hline
\rowcolor[HTML]{212121} 
{\color[HTML]{FFFFFF} Score} & {\color[HTML]{FFFFFF} Prepared} & {\color[HTML]{FFFFFF} Positive} & {\color[HTML]{FFFFFF} \begin{tabular}[c]{@{}c@{}}Hold High \\ Expectations\end{tabular}} & {\color[HTML]{FFFFFF} Creative} & {\color[HTML]{FFFFFF} Fair} & {\color[HTML]{FFFFFF} \begin{tabular}[c]{@{}c@{}}Display a \\ Personal Touch\end{tabular}} & {\color[HTML]{FFFFFF} \begin{tabular}[c]{@{}c@{}}Cultivate a Sense \\ of Belonging\end{tabular}} & {\color[HTML]{FFFFFF} Compassionate} & {\color[HTML]{FFFFFF} \begin{tabular}[c]{@{}c@{}}Have a Sense \\ of Humour\end{tabular}} & {\color[HTML]{FFFFFF} \begin{tabular}[c]{@{}c@{}}Respect \\ Students\end{tabular}} & {\color[HTML]{FFFFFF} Forgiving} & {\color[HTML]{FFFFFF} \begin{tabular}[c]{@{}c@{}}Admit \\ Mistakes\end{tabular}} \\ \hline
-2(MathDial Better)          & 27                              & 13                              & 11                                                                                       & 23                              & 8                           & 63                                                                                         & 21                                                                                               & 19                                   & 27                                                                                       & 5                                                                                  & 4                                & 30                                                                               \\
-1                           & 1                               & 0                               & 0                                                                                        & 9                               & 0                           & 8                                                                                          & 0                                                                                                & 0                                    & 31                                                                                       & 5                                                                                  & 4                                & 21                                                                               \\
0(Both Equal)                & 30                              & 25                              & 37                                                                                       & 36                              & 33                          & 38                                                                                         & 30                                                                                               & 39                                   & 79                                                                                       & 40                                                                                 & 32                               & 47                                                                               \\
1                            & 0                               & 0                               & 0                                                                                        & 24                              & 11                          & 19                                                                                         & 0                                                                                                & 0                                    & 49                                                                                       & 14                                                                                 & 19                               & 29                                                                               \\
2(MWPTutor Better)           & 152                             & 172                             & 162                                                                                      & 118                             & 158                         & 82                                                                                         & 159                                                                                              & 152                                  & 24                                                                                       & 146                                                                                & 151                              & 83                                                                               \\ \hline
\end{tabular}}
\caption{GPT Evaluations of metrics from Walker}
\label{tab:Walker_met}
\end{table*}

%% file: tables/all_corr.tex
% \usepackage{multirow}
\begin{table*}[]
\resizebox{\linewidth}{!}{\begin{tabular}{|c|c|c|c|c|}
\hline
\rowcolor[HTML]{212121} 
{\color[HTML]{FFFFFF} Source}   & {\color[HTML]{FFFFFF} Index} & {\color[HTML]{FFFFFF} Metric}                                                                             & {\color[HTML]{FFFFFF} \begin{tabular}[c]{@{}c@{}}Applicable to \\ Our Setting\end{tabular}} & {\color[HTML]{FFFFFF} Corresponding Metric} \\ \hline
                                & 1.1                          & Mistake Identification                                                                                    & Yes                                                                                         & Engagement                                  \\ \cline{2-5} 
                                & 1.2                          & Mistake Location                                                                                          & Yes                                                                                         & Engagement                                  \\ \cline{2-5} 
                                & 1.3                          & Revealing The Answer                                                                                      & Partially                                                                                   & Scaffolding                                 \\ \cline{2-5} 
                                & 1.4                          & Providing Guidance                                                                                        & Yes                                                                                         & Scaffolding                                 \\ \cline{2-5} 
                                & 1.5                          & Actionability                                                                                             & Yes                                                                                         & Engagement                                  \\ \cline{2-5} 
                                & 1.6                          & Coherence                                                                                                 & Yes                                                                                         & Engagement                                  \\ \cline{2-5} 
                                & 1.7                          & Tutor tone                                                                                                & Yes                                                                                         & Empathy                                     \\ \cline{2-5} 
\multirow{-8}{*}{Maurya et al.\cite{maurya2024unifying}} & 1.8                          & Human Likeness                                                                                            & Yes                                                                                         & Empathy                                     \\ \hline
                                & 2.1                          & Prepared                                                                                                  & Partially                                                                                   & Engagement                                  \\ \cline{2-5} 
                                & 2.2                          & Positive                                                                                                  & Yes                                                                                         & Empathy                                     \\ \cline{2-5} 
                                & 2.3                          & Hold High Expectations                                                                                    & No                                                                                          & N/A                                         \\ \cline{2-5} 
                                & 2.4                          & Creative                                                                                                  & No                                                                                          & N/A                                         \\ \cline{2-5} 
                                & 2.5                          & Fair                                                                                                      & Partially                                                                                   & Empathy                                     \\ \cline{2-5} 
                                & 2.6                          & Display a Personal Touch                                                                                  & Partially                                                                                   & Empathy                                     \\ \cline{2-5} 
                                & 2.7                          & Cultivate a Sense of Belonging                                                                            & No                                                                                          & N/A                                         \\ \cline{2-5} 
                                & 2.8                          & Compassionate                                                                                             & Yes                                                                                         & Empathy                                     \\ \cline{2-5} 
                                & 2.9                          & Have a Sense of Humour                                                                                    & Yes                                                                                         & N/A                                         \\ \cline{2-5} 
                                & 2.10                         & Respect Students                                                                                          & Yes                                                                                         & Empathy                                     \\ \cline{2-5} 
                                & 2.11                         & Forgiving                                                                                                 & Partially                                                                                   & Empathy                                     \\ \cline{2-5} 
\multirow{-12}{*}{Walker\cite{walker2008twelve}}       & 2.12                         & Admit Mistakes                                                                                            & No                                                                                          & N/A                                         \\ \hline
                                & 3.1                          & Promote independence in learning                                                                          & Yes                                                                                         & Scaffolding                                 \\ \cline{2-5} 
                                & 3.2                          & Personalize instruction                                                                                   & Partially                                                                                   & Engagement                                  \\ \cline{2-5} 
                                & 3.3                          & \begin{tabular}[c]{@{}c@{}}Facilitate tutee insights into\\  learning and learning processes\end{tabular} & Yes                                                                                         & Scaffolding                                 \\ \cline{2-5} 
                                & 3.4                          & \begin{tabular}[c]{@{}c@{}}Provide a student perspective on \\ learning and school success\end{tabular}   & No                                                                                          & N/A                                         \\ \cline{2-5} 
                                & 3.5                          & Respect individual differences                                                                            & Partially                                                                                   & Empathy                                     \\ \cline{2-5} 
\multirow{-6}{*}{MacDonald\cite{macdonald2000master}}     & 3.6                          & Follow a Job Description                                                                                  & No                                                                                          & N/A                                         \\ \hline
\end{tabular}}
\caption{List of Metrics defined by related work and their mapping to corresponding metrics used by us. We refer interested readers to the original works for full definitions of the metrics. We number the metrics to make it easier for us to refer to them in text. }
\label{tab:Corr}
\end{table*}

%% file: tables/ann_wise.tex
% \usepackage{multirow}
\begin{table*}[]
\begin{tabular}{|cc|cc|cc|cc|cc|}
\hline
\rowcolor[HTML]{333333} 
\multicolumn{1}{|c|}{\cellcolor[HTML]{333333}{\color[HTML]{FFFFFF} }}                                                                              & \cellcolor[HTML]{333333}{\color[HTML]{FFFFFF} }                                                                                         & \multicolumn{2}{c|}{\cellcolor[HTML]{333333}{\color[HTML]{FFFFFF} \textbf{Conciseness}}}                                                                                                                                      & \multicolumn{2}{c|}{\cellcolor[HTML]{333333}{\color[HTML]{FFFFFF} \textbf{Engagement}}}                                                                                       & \multicolumn{2}{c|}{\cellcolor[HTML]{333333}{\color[HTML]{FFFFFF} \textbf{Empathy}}}                                                                                                                                       & \multicolumn{2}{c|}{\cellcolor[HTML]{333333}{\color[HTML]{FFFFFF} \textbf{Scaffolding}}}                                                                                                                                   \\ \cline{3-10} 
\rowcolor[HTML]{333333} 
\multicolumn{1}{|c|}{\multirow{-2}{*}{\cellcolor[HTML]{333333}{\color[HTML]{FFFFFF} \textbf{\begin{tabular}[c]{@{}c@{}}Ann.\\ No.\end{tabular}}}}} & \multirow{-2}{*}{\cellcolor[HTML]{333333}{\color[HTML]{FFFFFF} \textbf{\begin{tabular}[c]{@{}c@{}}Questions\\ Annotated\end{tabular}}}} & \cellcolor[HTML]{333333}{\color[HTML]{FFFFFF} \textbf{\begin{tabular}[c]{@{}c@{}}LLM \\ Better\end{tabular}}} & \cellcolor[HTML]{333333}{\color[HTML]{FFFFFF} \textbf{\begin{tabular}[c]{@{}c@{}}Both\\  Equal\end{tabular}}} & {\color[HTML]{FFFFFF} \textbf{\begin{tabular}[c]{@{}c@{}}LLM \\ Better\end{tabular}}} & {\color[HTML]{FFFFFF} \textbf{\begin{tabular}[c]{@{}c@{}}Both\\  Equal\end{tabular}}} & \multicolumn{1}{c|}{\cellcolor[HTML]{333333}{\color[HTML]{FFFFFF} \textbf{\begin{tabular}[c]{@{}c@{}}LLM \\ Better\end{tabular}}}} & {\color[HTML]{FFFFFF} \textbf{\begin{tabular}[c]{@{}c@{}}Both\\  Equal\end{tabular}}} & \multicolumn{1}{c|}{\cellcolor[HTML]{333333}{\color[HTML]{FFFFFF} \textbf{\begin{tabular}[c]{@{}c@{}}LLM \\ Better\end{tabular}}}} & {\color[HTML]{FFFFFF} \textbf{\begin{tabular}[c]{@{}c@{}}Both\\  Equal\end{tabular}}} \\ \hline
\multicolumn{1}{|c|}{1}                                                                                                                            & 1-30                                                                                                                                    & \textbf{12}                                                                                                   & \textbf{0}                                                                                                    & 19                                                                                    & 0                                                                                     & \multicolumn{1}{c|}{15}                                                                                                            & 0                                                                                     & \multicolumn{1}{c|}{27}                                                                                                            & 0                                                                                     \\
\multicolumn{1}{|c|}{2}                                                                                                                            & 1-30                                                                                                                                    & 16                                                                                                            & 1                                                                                                             & 18                                                                                    & 0                                                                                     & \multicolumn{1}{c|}{17}                                                                                                            & 1                                                                                     & \multicolumn{1}{c|}{19}                                                                                                            & 1                                                                                     \\
\multicolumn{1}{|c|}{3}                                                                                                                            & 1-30                                                                                                                                    & \textbf{10}                                                                                                   & \textbf{2}                                                                                                    & \textbf{13}                                                                           & \textbf{0}                                                                            & \multicolumn{1}{c|}{15}                                                                                                            & 0                                                                                     & \multicolumn{1}{c|}{15}                                                                                                            & 0                                                                                     \\
\multicolumn{1}{|c|}{4}                                                                                                                            & 1-30                                                                                                                                    & \textbf{13}                                                                                                   & \textbf{0}                                                                                                    & \textbf{11}                                                                           & \textbf{0}                                                                            & \multicolumn{1}{c|}{17}                                                                                                            & 0                                                                                     & \multicolumn{1}{c|}{\textbf{12}}                                                                                                   & \textbf{0}                                                                            \\
\multicolumn{1}{|c|}{5}                                                                                                                            & 1-30                                                                                                                                    & 16                                                                                                            & 0                                                                                                             & 18                                                                                    & 0                                                                                     & \multicolumn{1}{c|}{15}                                                                                                            & 0                                                                                     & \multicolumn{1}{c|}{16}                                                                                                            & 0                                                                                     \\
\multicolumn{1}{|c|}{6}                                                                                                                            & 31-60                                                                                                                                   & \textbf{14}                                                                                                   & \textbf{1}                                                                                                    & \textbf{14}                                                                           & \textbf{1}                                                                            & \multicolumn{1}{c|}{16}                                                                                                            & 2                                                                                     & \multicolumn{1}{c|}{\textbf{14}}                                                                                                   & \textbf{1}                                                                            \\
\multicolumn{1}{|c|}{7}                                                                                                                            & 31-60                                                                                                                                   & 16                                                                                                            & 1                                                                                                             & 17                                                                                    & 0                                                                                     & \multicolumn{1}{c|}{\textbf{14}}                                                                                                   & \textbf{0}                                                                            & \multicolumn{1}{c|}{17}                                                                                                            & 0                                                                                     \\
\multicolumn{1}{|c|}{8}                                                                                                                            & 31-60                                                                                                                                   & 23                                                                                                            & 0                                                                                                             & 23                                                                                    & 2                                                                                     & \multicolumn{1}{c|}{24}                                                                                                            & 2                                                                                     & \multicolumn{1}{c|}{21}                                                                                                            & 2                                                                                     \\
\multicolumn{1}{|c|}{9}                                                                                                                            & 31-60                                                                                                                                   & \textbf{13}                                                                                                   & \textbf{1}                                                                                                    & 14                                                                                    & 5                                                                                     & \multicolumn{1}{c|}{20}                                                                                                            & 5                                                                                     & \multicolumn{1}{c|}{16}                                                                                                            & 4                                                                                     \\
\multicolumn{1}{|c|}{10}                                                                                                                           & 31-60                                                                                                                                   & 18                                                                                                            & 7                                                                                                             & 16                                                                                    & 0                                                                                     & \multicolumn{1}{c|}{16}                                                                                                            & 5                                                                                     & \multicolumn{1}{c|}{16}                                                                                                            & 2                                                                                     \\
\multicolumn{1}{|c|}{11}                                                                                                                           & 61-90                                                                                                                                   & 17                                                                                                            & 1                                                                                                             & 14                                                                                    & 2                                                                                     & \multicolumn{1}{c|}{16}                                                                                                            & 4                                                                                     & \multicolumn{1}{c|}{\textbf{11}}                                                                                                   & \textbf{4}                                                                            \\
\multicolumn{1}{|c|}{12}                                                                                                                           & 61-90                                                                                                                                   & 15                                                                                                            & 0                                                                                                             & \textbf{8}                                                                            & \textbf{0}                                                                            & \multicolumn{1}{c|}{\textbf{12}}                                                                                                   & \textbf{0}                                                                            & \multicolumn{1}{c|}{\textbf{13}}                                                                                                   & \textbf{0}                                                                            \\
\multicolumn{1}{|c|}{13}                                                                                                                           & 61-90                                                                                                                                   & 21                                                                                                            & 1                                                                                                             & 24                                                                                    & 0                                                                                     & \multicolumn{1}{c|}{24}                                                                                                            & 0                                                                                     & \multicolumn{1}{c|}{24}                                                                                                            & 0                                                                                     \\
\multicolumn{1}{|c|}{14}                                                                                                                           & 61-90                                                                                                                                   & 10                                                                                                            & 11                                                                                                            & 13                                                                                    & 11                                                                                    & \multicolumn{1}{c|}{11}                                                                                                            & 11                                                                                    & \multicolumn{1}{c|}{\textbf{9}}                                                                                                    & \textbf{10}                                                                           \\
\multicolumn{1}{|c|}{15}                                                                                                                           & 61-90                                                                                                                                   & 20                                                                                                            & 4                                                                                                             & 20                                                                                    & 2                                                                                     & \multicolumn{1}{c|}{16}                                                                                                            & 5                                                                                     & \multicolumn{1}{c|}{18}                                                                                                            & 2                                                                                     \\
\multicolumn{1}{|c|}{16}                                                                                                                           & 91-120                                                                                                                                  & 20                                                                                                            & 1                                                                                                             & 20                                                                                    & 0                                                                                     & \multicolumn{1}{c|}{22}                                                                                                            & 0                                                                                     & \multicolumn{1}{c|}{20}                                                                                                            & 1                                                                                     \\
\multicolumn{1}{|c|}{17}                                                                                                                           & 91-120                                                                                                                                  & 15                                                                                                            & 1                                                                                                             & \textbf{12}                                                                           & \textbf{2}                                                                            & \multicolumn{1}{c|}{14}                                                                                                            & 2                                                                                     & \multicolumn{1}{c|}{15}                                                                                                            & 1                                                                                     \\
\multicolumn{1}{|c|}{18}                                                                                                                           & 91-120                                                                                                                                  & \textbf{13}                                                                                                   & \textbf{2}                                                                                                    & \textbf{11}                                                                           & \textbf{0}                                                                            & \multicolumn{1}{c|}{14}                                                                                                            & 3                                                                                     & \multicolumn{1}{c|}{21}                                                                                                            & 0                                                                                     \\
\multicolumn{1}{|c|}{19}                                                                                                                           & 91-120                                                                                                                                  & 15                                                                                                            & 0                                                                                                             & 17                                                                                    & 0                                                                                     & \multicolumn{1}{c|}{\textbf{12}}                                                                                                   & \textbf{0}                                                                            & \multicolumn{1}{c|}{\textbf{14}}                                                                                                   & \textbf{0}                                                                            \\
\multicolumn{1}{|c|}{20}                                                                                                                           & 91-120                                                                                                                                  & \textbf{12}                                                                                                   & \textbf{3}                                                                                                    & \textbf{12}                                                                           & \textbf{0}                                                                            & \multicolumn{1}{c|}{15}                                                                                                            & 1                                                                                     & \multicolumn{1}{c|}{\textbf{12}}                                                                                                   & \textbf{2}                                                                            \\
\multicolumn{1}{|c|}{21}                                                                                                                           & 121-150                                                                                                                                 & 11                                                                                                            & 14                                                                                                            & 6                                                                                     & 21                                                                                    & \multicolumn{1}{c|}{12}                                                                                                            & 15                                                                                    & \multicolumn{1}{c|}{15}                                                                                                            & 8                                                                                     \\
\multicolumn{1}{|c|}{22}                                                                                                                           & 121-150                                                                                                                                 & 14                                                                                                            & 3                                                                                                             & \textbf{11}                                                                           & \textbf{6}                                                                            & \multicolumn{1}{c|}{13}                                                                                                            & 6                                                                                     & \multicolumn{1}{c|}{\textbf{13}}                                                                                                   & \textbf{2}                                                                            \\
\multicolumn{1}{|c|}{23}                                                                                                                           & 121-150                                                                                                                                 & 15                                                                                                            & 5                                                                                                             & 15                                                                                    & 2                                                                                     & \multicolumn{1}{c|}{13}                                                                                                            & 9                                                                                     & \multicolumn{1}{c|}{\textbf{11}}                                                                                                   & \textbf{7}                                                                            \\
\multicolumn{1}{|c|}{24}                                                                                                                           & 121-150                                                                                                                                 & 12                                                                                                            & 11                                                                                                            & 12                                                                                    & 9                                                                                     & \multicolumn{1}{c|}{12}                                                                                                            & 12                                                                                    & \multicolumn{1}{c|}{13}                                                                                                            & 9                                                                                     \\
\multicolumn{1}{|c|}{25}                                                                                                                           & 121-150                                                                                                                                 & 17                                                                                                            & 0                                                                                                             & \textbf{11}                                                                           & \textbf{0}                                                                            & \multicolumn{1}{c|}{15}                                                                                                            & 0                                                                                     & \multicolumn{1}{c|}{\textbf{13}}                                                                                                   & \textbf{1}                                                                            \\
\multicolumn{1}{|c|}{26}                                                                                                                           & 151-180                                                                                                                                 & 27                                                                                                            & 0                                                                                                             & 16                                                                                    & 0                                                                                     & \multicolumn{1}{c|}{22}                                                                                                            & 0                                                                                     & \multicolumn{1}{c|}{21}                                                                                                            & 0                                                                                     \\
\multicolumn{1}{|c|}{27}                                                                                                                           & 151-180                                                                                                                                 & 17                                                                                                            & 0                                                                                                             & \textbf{13}                                                                           & \textbf{1}                                                                            & \multicolumn{1}{c|}{\textbf{6}}                                                                                                    & \textbf{9}                                                                            & \multicolumn{1}{c|}{\textbf{9}}                                                                                                    & \textbf{10}                                                                           \\
\multicolumn{1}{|c|}{28}                                                                                                                           & 151-180                                                                                                                                 & \textbf{12}                                                                                                   & \textbf{1}                                                                                                    & 20                                                                                    & 0                                                                                     & \multicolumn{1}{c|}{17}                                                                                                            & 0                                                                                     & \multicolumn{1}{c|}{15}                                                                                                            & 0                                                                                     \\
\multicolumn{1}{|c|}{29}                                                                                                                           & 151-180                                                                                                                                 & 17                                                                                                            & 2                                                                                                             & 16                                                                                    & 1                                                                                     & \multicolumn{1}{c|}{\textbf{14}}                                                                                                   & \textbf{1}                                                                            & \multicolumn{1}{c|}{14}                                                                                                            & 3                                                                                     \\
\multicolumn{1}{|c|}{30}                                                                                                                           & 151-180                                                                                                                                 & 21                                                                                                            & 1                                                                                                             & 21                                                                                    & 1                                                                                     & \multicolumn{1}{c|}{23}                                                                                                            & 1                                                                                     & \multicolumn{1}{c|}{22}                                                                                                            & 1                                                                                     \\
\multicolumn{1}{|c|}{31}                                                                                                                           & 181-210                                                                                                                                 & 17                                                                                                            & 0                                                                                                             & \textbf{12}                                                                           & \textbf{0}                                                                            & \multicolumn{1}{c|}{16}                                                                                                            & 1                                                                                     & \multicolumn{1}{c|}{16}                                                                                                            & 0                                                                                     \\
\multicolumn{1}{|c|}{32}                                                                                                                           & 181-210                                                                                                                                 & \textbf{13}                                                                                                   & \textbf{1}                                                                                                    & \textbf{11}                                                                           & \textbf{1}                                                                            & \multicolumn{1}{c|}{\textbf{9}}                                                                                                    & \textbf{4}                                                                            & \multicolumn{1}{c|}{\textbf{12}}                                                                                                   & \textbf{2}                                                                            \\
\multicolumn{1}{|c|}{33}                                                                                                                           & 181-210                                                                                                                                 & 16                                                                                                            & 0                                                                                                             & \textbf{14}                                                                           & \textbf{0}                                                                            & \multicolumn{1}{c|}{14}                                                                                                            & 3                                                                                     & \multicolumn{1}{c|}{\textbf{14}}                                                                                                   & \textbf{0}                                                                            \\
\multicolumn{1}{|c|}{34}                                                                                                                           & 181-210                                                                                                                                 & \textbf{11}                                                                                                   & \textbf{1}                                                                                                    & \textbf{10}                                                                           & \textbf{0}                                                                            & \multicolumn{1}{c|}{\textbf{9}}                                                                                                    & \textbf{6}                                                                            & \multicolumn{1}{c|}{\textbf{10}}                                                                                                   & \textbf{1}                                                                            \\
\multicolumn{1}{|c|}{35}                                                                                                                           & 181-210                                                                                                                                 & 12                                                                                                            & 8                                                                                                             & 5                                                                                     & 21                                                                                    & \multicolumn{1}{c|}{12}                                                                                                            & 15                                                                                    & \multicolumn{1}{c|}{7}                                                                                                             & 21                                                                                    \\ \hline
\multicolumn{2}{|c|}{\textbf{\begin{tabular}[c]{@{}c@{}}\% Not Favouring\\  Humans\end{tabular}}}                                                                                                                                                                                            & \multicolumn{2}{c|}{\textbf{71\%}}                                                                                                                                                                                            & \multicolumn{2}{c|}{\textbf{60\%}}                                                                                                                                            & \multicolumn{2}{c|}{\textbf{80\%}}                                                                                                                                                                                         & \multicolumn{2}{c|}{\textbf{60\%}}                                                                                                                                                                                         \\ \hline
\end{tabular}
\caption{Annotator-wise choice summary. Entries where annotator leans in favour of human (ie LLMBetter+0.5*Bothequal\textless{}15) are in bold. }
\label{tab:ann_wise}
\end{table*}